\documentclass[sigconf, screen]{acmart}
\usepackage{pifont}
\usepackage{multirow}
\usepackage{subcaption}
\usepackage{hyperref}
\hypersetup{
  colorlinks = true,
  linkcolor  = blue,
  citecolor  = blue,
  urlcolor   = blue
}
\makeatletter
\renewcommand\acmBadgeR[2][]{%
  \ifx\@acmBadgeR\@empty
    \gdef\@acmBadgeR{%
      \smash{\raisebox{-2pt}{\href{#1}{\includegraphics[width=\@ACM@badge@width]{#2}}}}}%
  \else
    \g@addto@macro{\@acmBadgeR}{%
      \hspace{\@ACM@badge@skip}%
      \smash{\raisebox{-2pt}{\href{#1}{\includegraphics[width=\@ACM@badge@width]{#2}}}}}%
  \fi
}
\makeatother

\AtBeginDocument{%
  }
\copyrightyear{2026}
\acmYear{2026}
\setcopyright{cc}
\setcctype{by-nc-nd}
\acmConference[ICSE '26]{2026 IEEE/ACM 48th International Conference on Software Engineering}{April 12--18, 2026}{Rio de Janeiro, Brazil}
\acmBooktitle{2026 IEEE/ACM 48th International Conference on Software Engineering (ICSE '26), April 12--18, 2026, Rio de Janeiro, Brazil}
\acmPrice{}
\acmDOI{10.1145/3744916.3787795}
\acmISBN{979-8-4007-2025-3/2026/04}

\acmBadgeR[https://www.acm.org/publications/policies/artifact-review-and-badging-current]{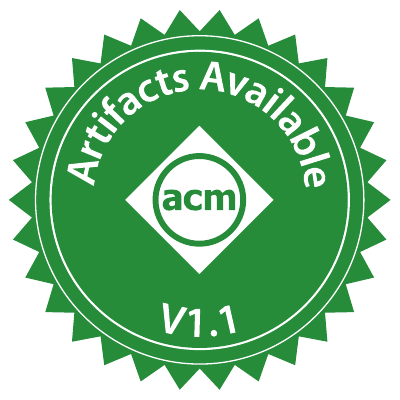}
\acmBadgeR[https://www.acm.org/publications/policies/artifact-review-and-badging-current]{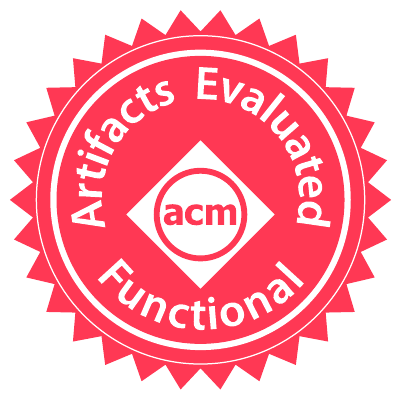}
\acmBadgeR[https://www.acm.org/publications/policies/artifact-review-and-badging-current]{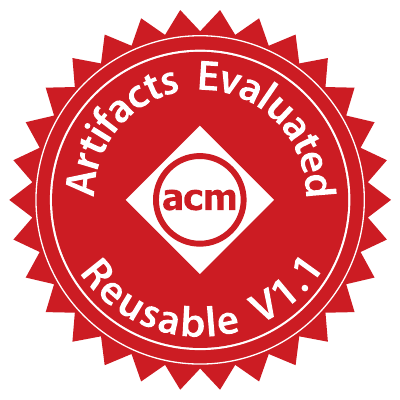}

\begin{document}

\title{Imitation Game: Reproducing Deep Learning Bugs Leveraging an Intelligent Agent}

\author{\href{https://orcid.org/0009-0000-3584-0470}{Mehil B Shah}\orcid{0009-0000-3584-0470}}\orcid{0009-0000-3584-0470}
\affiliation{%
  \institution{Dalhousie University}
  \city{Halifax}
  \country{Canada}}
\email{shahmehil@dal.ca}

\author{Mohammad Masudur Rahman}
\orcid{0000-0003-3821-5990}
\affiliation{%
  \institution{Dalhousie University}
  \city{Halifax}
  \country{Canada}}
\email{masud.rahman@dal.ca}

\author{Foutse Khomh}
\orcid{0000-0002-5704-4173}
\affiliation{%
 \institution{Polytechnique Montreal}
 \city{Montreal}
 \country{Canada}}
\email{foutse.khomh@polymtl.ca}

\begin{abstract}
Despite their wide adoption in various domains (e.g., healthcare, finance, software engineering), Deep Learning (DL)-based applications suffer from many bugs, failures, and vulnerabilities. Reproducing these bugs is essential for their resolution, but it is extremely challenging due to the inherent nondeterminism of DL models and their tight coupling with hardware and software environments. According to recent studies, only about 3\% of DL bugs can be reliably reproduced using manual approaches. To address these challenges, we present RepGen, a novel, automated, and intelligent approach for reproducing deep learning bugs. RepGen constructs a learning-enhanced context from a project, develops a comprehensive plan for bug reproduction, employs an iterative generate-validate-refine mechanism, and thus generates such code using an LLM that reproduces the bug at hand. We evaluate RepGen on 106 real-world deep learning bugs and achieve a reproduction rate of 80.19\%, a 19.81\% improvement over the state-of-the-art measure. A developer study involving 27 participants shows that RepGen improves the success rate of DL bug reproduction by 23.35\%, reduces the time to reproduce by 56.8\%, and lowers participants' cognitive load.
\end{abstract}



\maketitle
\vspace{-0.5em}

\section{Introduction}
Artificial Intelligence (AI) has been widely adopted in many application domains, including software engineering~\cite{mahbub2023explaining, mahbub2024predicting}, autonomous vehicles~\cite{grigorescu2020survey}, healthcare~\cite{med1}, finance~\cite{fin1}, and cybersecurity~\cite{cyb1}. The global market share of AI software reached \$34.8 billion in 2023 and is projected to grow up to \$360 billion by 2030~\cite{grandview2024}. Over 67\% of top-performing companies have incorporated AI in their business solutions, and 97\% of Fortune 500 companies have invested in AI technologies~\cite{pwc2023}, indicating their significance. However, software applications empowered by Deep Learning (DL), the underlying technology behind current AI systems, remain prone to bugs, faults, and vulnerabilities, which could lead to major consequences (e.g., system crashes) and catastrophic failures (e.g., autonomous vehicle accidents)~\cite{selfdrivingcarcrash}. Unlike the bugs in traditional, developer-written software, the bugs in DL software are inherently challenging due to several factors. First, they are often non-deterministic due to randomness in model training, i.e., random weight initialization of the model layers~\cite{nagarajan2018impact}. Second, DL models perform high-dimensional tensor operations and suffer from a lack of interpretability, making their encountered bugs opaque~\cite{mahbub2023explaining}. Finally, these bugs also have multi-faceted dependencies on hardware (e.g., GPU), underlying frameworks (e.g., PyTorch, TensorFlow)~\cite{tensorflowprogrambugs}, and data pipelines, making them highly complex~\cite{shah2024towards}. 

To resolve DL bugs, software developers must first systematically reproduce them on their local machines. Without a reproduction, they cannot confirm the presence of a bug or diagnose its root cause. However, reproduction of DL bugs can be effort-intensive, time-consuming, and frustrating due to various technical challenges. They include intricate data pipelines, hardware dependencies, and variations in software frameworks and library versions. Even when a bug is reproducible, developers frequently need to engage in trial-and-error, carefully tune environmental settings, and reason about the contextual factors that may influence the behaviour of DL programs, all of which can be tedious and error-prone~\cite{naziri2025bugsindlls}. Developers also face the challenge of missing or incomplete information when attempting to reproduce bugs from issue reports~\cite{wang2024systematic}. Reports may lack crucial details of a bug and omit relevant data or code snippets. In such cases, even experienced developers must spend substantial time reconstructing the missing detail of a bug, the target environment, and iteratively testing hypotheses to reproduce the erroneous behaviour~\cite{huang2023demystifying, zhang2020detecting}. These difficulties make the manual process of reproducing DL bugs highly inefficient and error-prone. Thus, there is a strong need for techniques that can \textit{accelerate and systematize} the reproduction of DL bugs, reducing developer effort while increasing the reliability of DL systems.

Over the last two decades, there have been many techniques to support developers in localizing~\cite{saha2013improving, mahmud2024using, chakraborty2024rlocator}, reproducing~\cite{kang2023large, nayrolles2015jcharming}, and correcting software bugs~\cite{bouzenia2024repairagent, li2022dear, xia2023automated}. However, they are not sufficient since they were not designed to tackle deep learning specific challenges. They might fail to effectively reproduce DL bugs or may do so inefficiently due to the following limitations:

\looseness=-1
\textbf{(a) Limited contextual understanding:} Existing techniques (e.g., RecDroid~\cite{zhao2019recdroid}, ReBL~\cite{wang2024feedback}, and AdbGPT~\cite{feng2024prompting}) analyze bug reports and GUI applications and leverage UI interactions or code execution paths to reproduce software bugs. In contrast, DL bugs go beyond source code and can originate from multiple artifacts scattered across different parts of the ML pipeline, including model architectures, training datasets, preprocessing scripts, hyperparameters, underlying frameworks, and hardware (e.g., GPU)~\cite{shah2025towards}. Collecting and correlating information from these scattered artifacts systematically is challenging, as each may contribute in subtle ways to the manifestation of the bug. Existing techniques lack mechanisms to identify or utilize these DL-specific contextual clues, even when they are present in bug reports or code, which makes these approaches inefficient or insufficient for reliably reproducing DL bugs.

\textbf{(b) Heavy reliance on GUI-specific elements:} Many existing techniques—RecDroid~\cite{zhao2019recdroid}, ReBL~\cite{wang2024feedback}, AdbGPT~\cite{feng2024prompting}, ReActDroid~\cite{huang2025one}, LLMDroid~\cite{wang2025llmdroid}—capture well-defined states of GUI-based applications using record-and-replay methods~\cite{honarmand2014replay} and generate deterministic event sequences to reproduce software faults. However, DL bugs originate from various stochastic processes such as model training, hyperparameter optimization, and data pipeline execution, which do not involve any GUI interactions~\cite{shah2025towards}. Thus, existing techniques relying on GUI interactions (e.g., mouse clicks, or screen capture) are fundamentally misaligned with DL-based systems, which makes them less effective for reproducing DL bugs.

\textbf{(c) Challenges with ``silent'' bugs:} Traditional techniques often verify the reproduction of a software bug by checking for crashes or undesirable program states~\cite{nayrolles2015jcharming, soltani2018single}. However, DL bugs can manifest themselves in non-conventional ways, such as increased losses during model training or extremely poor performance during model inference. Bugs with such symptoms are called \textit{silent} bugs in literature~\cite{tambon2024silent}. According to a recent study~\cite{tambon2024silent}, 72.8\% of developers reported difficulty in resolving silent bugs, which highlights the persistent challenges of these bugs. Thus, existing techniques~\cite{nayrolles2015jcharming, soltani2018single} relying on crashes or state-specific symptoms might be inadequate for accurately reproducing silent bugs from DL systems.

Given these limitations of existing techniques, we present \textit{RepGen}, an automated, intelligent approach for reproducing DL bugs, with a specific focus on accelerating and systematizing their reproduction. Unlike the existing techniques, our approach leverages a \textit{learning-enhanced context} for code generation and approximates \textit{fault symptoms of the code using LLMs}, thereby improving the speed, consistency, and reliability of DL bug reproduction. First, our technique builds a learning-enhanced context against a DL bug by capturing relevant code containing model training loops and their dependencies from a project. Second, using the context, RepGen generates a comprehensive plan for bug reproduction, which captures the environment setup, model training mechanism, and inference pipeline. Finally, both the context and the plan are passed to the LLM-powered agent, which runs an iterative \textit{generate–validate–refine} cycle to generate code that can reproduce the bug at hand. Our agent not only evaluates the code using standard mechanisms (e.g., compiler feedback, static analysis, relevance feedback)~\cite{samir2025improvedirbasedbuglocalization, tymchuk2018jit, wang-etal-2022-compilable} but also refines the code and validates its fault symptoms using an LLM (e.g., Qwen2.5-Coder-7B) to enhance its effectiveness in reproducing DL bugs. To the best of our knowledge, RepGen is the first technique to offer a unified, intelligent solution to systematically reproduce DL bugs, making our work \textit{novel}.

We evaluated RepGen using a comprehensive dataset of 106 real-world deep learning bugs collected from 16 GitHub projects. Our technique successfully reproduced 85 (80.19\%) bugs and outperformed eight LLM-only baselines (e.g., GPT 4.1, Llama3, DeepSeek-R1, Qwen3, Qwen2.5 and their variants), indicating its superiority. A controlled developer study with 27 participants further validated our results, showing that participants can reproduce $\approx$23.35\% more bugs when assisted by RepGen and can save $\approx$56.8\% time during bug reproduction. All these findings suggest a strong potential of our technique for supporting the systematic and efficient reproduction of deep learning bugs.

We thus make the following contributions in this paper:

(a) A novel, intelligent technique - RepGen - to support the reproduction of deep learning bugs, emphasizing the acceleration and systematization of bug reproduction.

(b) A novel agentic workflow that builds a comprehensive, learning-enhanced context, develops a targeted reproduction plan, and leverages a feedback-driven reproduction agent that can generate high-quality code capable of reproducing DL bugs efficiently.

(c) A comprehensive experiment comprising an empirical evaluation of RepGen, an ablation study targeting 13 major components, comparisons with eight baselines, and a developer study involving 27 participants, demonstrating the benefits of our technique.

(d) A carefully curated benchmark dataset containing 106 bug reports and associated artifacts (e.g., code examples, configuration files) from 16 real-world deep learning projects.

(e) A replication package~\cite{replicationpackage} containing our implementation, dataset, and experimental results to enable future research.

\vspace{-0.5em}
\section{Motivating Example}
\begin{figure}[]
    \centering
    \captionsetup{font=small}
    \includegraphics[width=0.85\columnwidth]{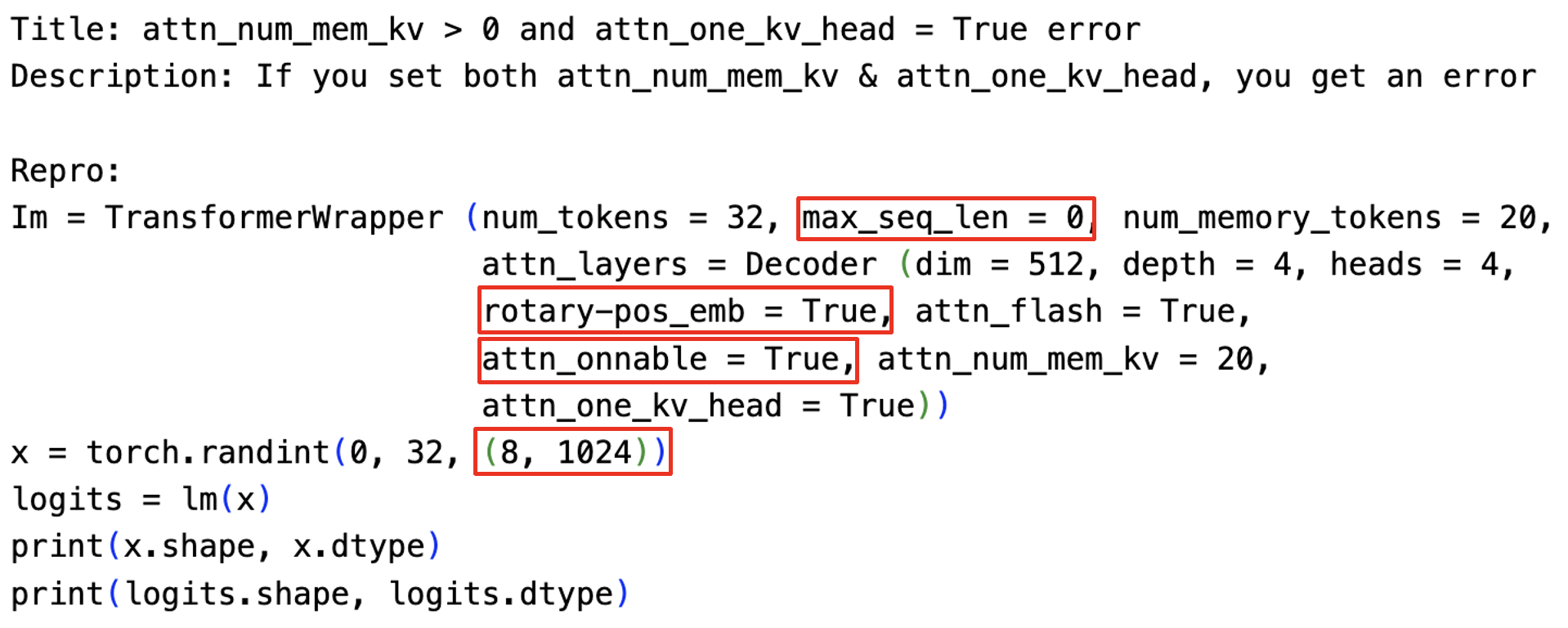}
    \vspace{-1.0em}
    \caption{Bug \#228 in X-Transformer; red boxes indicate incorrect or missing parameters in the user-provided snippet.}
    \label{fig:motivatingexample}
    \vspace{-1.5em}
\end{figure}

To demonstrate the effectiveness of our technique, let us consider the bug in Fig.~\ref{fig:motivatingexample} (Bug \#228, X-transformer). The bug occurs in the \texttt{TransformerWrapper} class when two specific parameters -- \texttt{attn\_num\_mem\_kv} and \texttt{attn\_one\_kv\_head} -- are initialized simultaneously. In particular, when the model uses a single shared key-value head (\texttt{attn\_one\_kv\_head = True}) and simultaneously allocates the memory for additional key-value pairs (\texttt{attn\_num\_mem\_kv = 20}), it leads to an incompatible configuration for the attention model, which causes a \texttt{ValueError}. 

As shown in Fig.~\ref{fig:motivatingexample}, the bug report provides a partially complete code snippet and lacks necessary details (e.g., required libraries). Reproducing such bugs manually is possible, but it requires developers to identify missing dependencies, understand parameter interdependencies, configure the target environment, and iteratively test for the reported faulty behaviour, which can be time-consuming and error-prone. Existing techniques for bug reproduction - RecD\-roid~\cite{zhao2019recdroid}, AdbGPT~\cite{feng2024prompting} - rely heavily on GUI-based interactions and event sequences, making them non-applicable to DL models operating through training data, model structure, hardware configurations, and underlying frameworks. Hence, they naturally fail to reproduce such bugs. On the other hand, ChatGPT, a high-performing LLM, accepts the bug report (Fig.~\ref{fig:motivatingexample}) and generates a non-compilable code snippet\footnote{http://bit.ly/462AOlj} with incorrect import statements.

\begin{figure}[!t]
    \centering
    \captionsetup{font=small}
    \includegraphics[width=0.85\columnwidth]{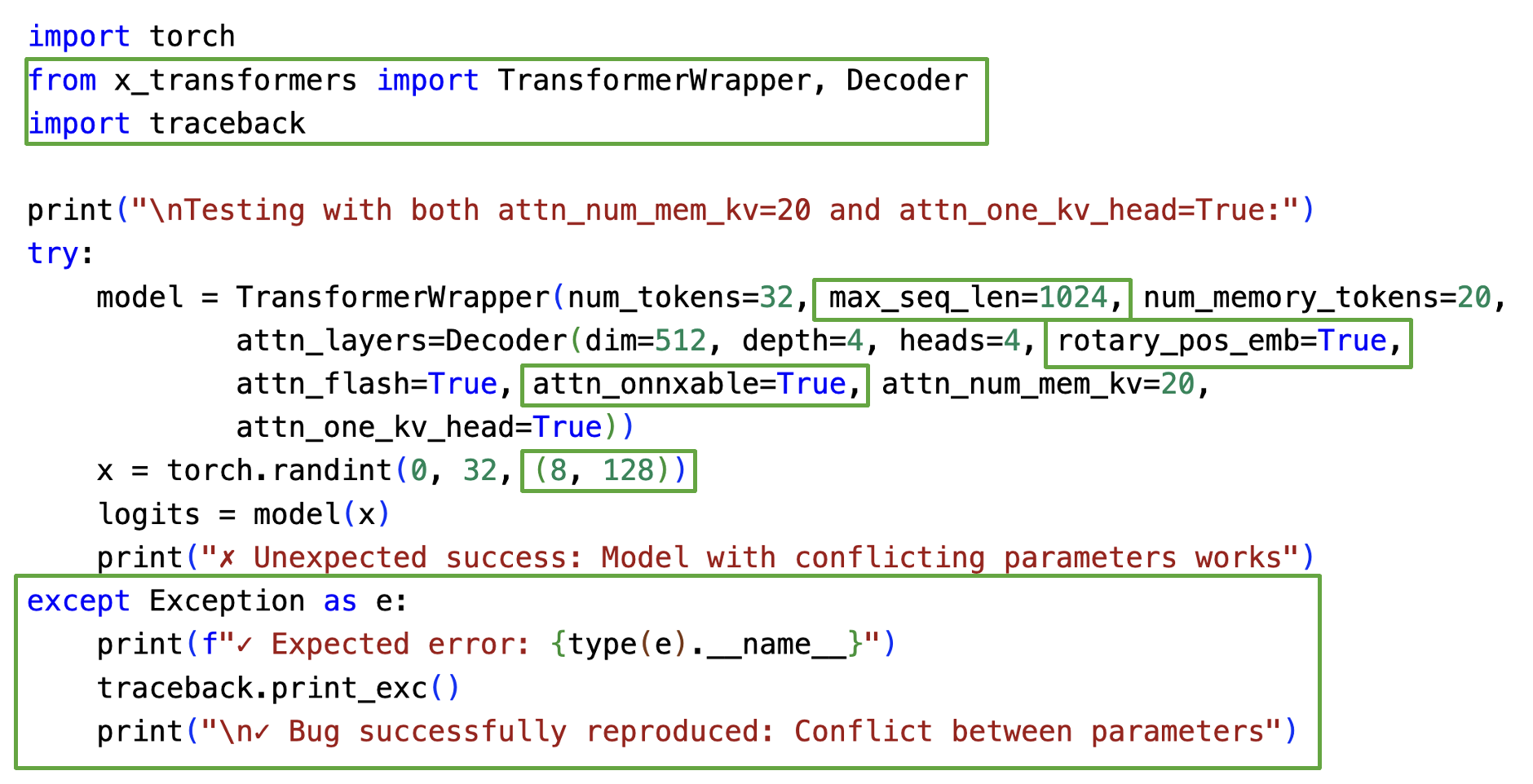}
    \vspace{-1.0em}
    \caption{Reproduction code generated by RepGen, with green boxes highlighting additions that make the script verifiable and executable.}
    \label{fig:repgen-output}
    \vspace{-1.5em}
\end{figure}

In contrast, our technique, RepGen, systematizes the bug reproduction process and provides a code example (Fig.~\ref{fig:repgen-output}) that successfully reproduces the bug. First, RepGen constructs a \textit{learning-enhanced context} by combining the code retrieved from the \texttt{Trans\-formerWrapper} and \texttt{Decoder} classes with the useful information from the bug report (e.g., erroneous behaviors). Second, RepGen leverages the context and formulates an appropriate plan to reproduce the bug. Finally, the context and the plan are passed to our reproduction agent. The reproduction agent then generates an initial version of code that misses required import statements, as detected by PyLint. Using PyLint's feedback, our reproduction agent generates a more complete and relevant code snippet, but it still lacks an important configuration needed to reproduce the bug (e.g., both \texttt{attn\_num\_mem\_kv} and \texttt{attn\_one\_kv\_head} should be initialized simultaneously). This issue was detected by our reproduction agent, suggesting that the code missed the symptoms mentioned in the bug report. By incorporating this feedback into the regeneration process, RepGen then generates a complete code example that can reproduce the bug (Fig.~\ref{fig:repgen-output}). 

\looseness=-1
In short, by leveraging a learning-enhanced context, comprehensive planning, multi-phase feedback, and iterative refinement, our technique enables the automated and structured reproduction of DL bugs that would otherwise require substantial manual effort. RepGen completes this process in $\approx$5 minutes, demonstrating its ability to accelerate and systematize bug reproduction. This also highlights our technique's potential for improving developer productivity and ensuring consistent, verifiable reproduction of DL bugs.

\vspace{-0.5em}
\section{RepGen}
\begin{figure*}
    \centering
    \includegraphics[page = 2, width=0.65\linewidth]{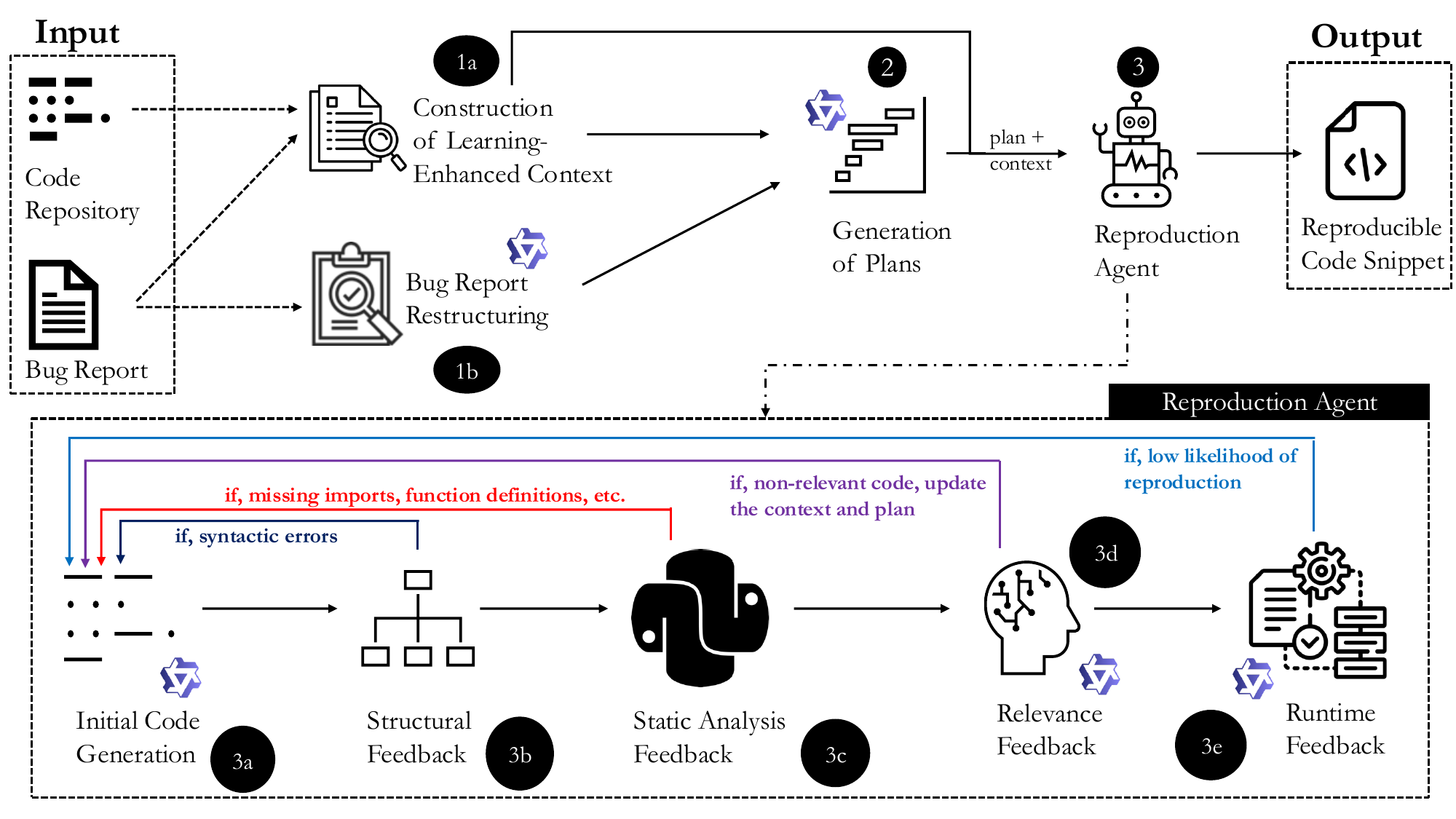}
    \vspace{-1.0em}
    \caption{Schematic diagram of our proposed technique - RepGen}
    \vspace{-1.5em}
    \label{fig:schematic}
\end{figure*}

Fig.~\ref{fig:schematic} shows the schematic diagram of our proposed technique -- RepGen-- for DL bug reproduction. We discuss the major steps of our methodology as follows.

\vspace{-0.3em}
\subsection{Construction of Learning-Enhanced Context}
\looseness=-1
Since we use an LLM to generate code capable of reproducing bugs from real-life projects, we need project-specific context. We thus leverage a hybrid retrieval framework and construct a learning-enhanced context against a bug report as follows (Step 1a, Fig.~\ref{fig:schematic}).

\subsubsection{Pre-Retrieval}
Before performing any retrieval, we apply a series of preprocessing steps to the codebase as follows.

\textbf{(a) Semantic chunking:} We break each source code file into smaller, semantically meaningful segments through code chunking~\cite{lewis2020retrieval}. It is essential since DL codebases often contain long, complex files spanning hundreds of lines. We employ Abstract Syntax Tree (AST) parsing, identify natural partitioning points (e.g., method boundary, class declaration), and break each file into smaller chunks that are structurally correct and semantically relevant.

\textbf{(b) Indexing:} After splitting the code into chunks, we index them for efficient search and real-time analysis. We construct two complementary indices for efficient retrieval of relevant code snippets: a sparse retrieval index that can be leveraged by the BM25 ranking function~\cite{robertson2009probabilistic} and a dense retrieval index that can be leveraged for embedding-based semantic retrieval.

\subsubsection{Code Snippet Retrieval}
To retrieve relevant code against a bug report, we design a hybrid retrieval approach leveraging the above indices as follows.

\textbf{(a) BM25 search:} Best Matching 25 (BM25) is a probabilistic algorithm for similarity matching that incorporates document length normalization and term frequency saturation. It can identify code chunks that contain exact matches of API calls, error messages, or specific identifiers found in a bug report. Given a query $Q$ derived from the bug description, we compute the BM25 score for a code chunk $D$~\cite{robertson2009probabilistic}. We also normalize the BM25 scores across all the chunks for an effective analysis~\cite{zhang2024ragenhancedcommitmessagegeneration}.

\looseness=-1
\textbf{(b) ANN search:} Approximate Nearest Neighbour (ANN) search leverages semantic proximity between two items within the semantic space. We use an ANN search to capture the conceptual or semantic relationship between code chunks and bug reports that may not be evident from their contents. Using a code-language model\footnote{http://bit.ly/3IwN0kk}, we encode each bug report and code chunk into dense vectors and normalize them. Then, to measure the similarity between the bug report (\textbf{q}) and code (\textbf{d}), we compute their angular similarity~\cite{angular}.

\textbf{(c) Hybrid scoring:} To leverage the complementary strengths of both approaches, we combine the normalized scores from the BM25 Search and ANN Search using the formula below:
\[
\text{score}_{\text{h}}(D) = (1 - \alpha) \cdot \text{BM25}_{\text{norm}}(Q, D) + \alpha \cdot \text{sim}_{\text{angular}}(\mathbf{q}, \mathbf{d})
\]

\noindent
After iterative hyperparameter tuning, we set $\alpha$ to 0.55. We then rank the code snippets by their hybrid scores and select the top K (e.g., 20) for further re-ranking.

\textbf{(d) Reranking:}
After retrieving a short list of code snippets above, we employ a cross-encoder\footnote{http://bit.ly/454NWFo} to re-rank them. The \textit{ms-marco-MiniLM-L12-v2} cross-encoder demonstrates a strong capability in understanding texts and code, making it a suitable choice for our approach. First, MiniLMv2, distilled from the BERT and RoBERTa architectures, inherits their strong performance on code-specific tasks~\cite{bert2021msr, wang2022bridging, karmakar2021pre}. Second, its fine-tuning on the MS Marco dataset~\cite{bajaj2016ms}, derived from Bing search queries, provides a comprehensive understanding of human-generated natural language. Unlike the above retrieval methods, which rely on lexical or semantic matching, the cross-encoder jointly encodes both the bug report and code snippet, allowing its attention mechanisms to detect their exact relationships. It can also identify subtle but important connections that the simpler retrieval methods might miss.

Given a bug report $Q$ and code snippet $D$, the cross-encoder computes a relevance score, as follows:
\vspace{-0.3em}
\[
\text{score}_{\text{cross}}(Q,D) = \text{CrossEncoder}([\text{CLS}] \oplus Q \oplus [\text{SEP}] \oplus D \oplus [\text{SEP}])
\]
\noindent
where $\oplus$ denotes concatenation with special tokens (e.g., CLS, SEP), and the model analyzes cross-attention between all tokens to produce a score between 0 and 1. Based on the cross-encoder scores, we re-rank the top K snippets (e.g., 20) for further processing. 

\textbf{(e) Capturing dependencies:} Deep learning pipeline involves multiple components, and their bugs could be multifaceted. For example, a bug in a model training loop might depend on the custom layer's definition, data preprocessing functions, and custom loss functions defined in separate modules. Therefore, for each retrieved snippet $D_i$, we parse its AST and identify the corresponding dependencies (e.g., imported modules, referenced variables). The dependency resolution ensures that all necessary code components are captured for subsequent analysis.

\subsubsection{Context Construction}
Once relevant code snippets and their dependencies are captured, we construct a unified context that can help LLMs generate appropriate code reproducing deep learning bugs. To achieve this, we perform several analyses using the retrieved code and bug report as follows.

\textbf{(a) Module-centric partitioning \& retrieval:} 
We organize the retrieved code snippets based on their source modules. Our idea was to select the modules containing the most relevant code to the bug. We thus map the code snippets to their corresponding system paths and organize them in modules:
\[
M_k = \text{ResolveModule}(D_i), \quad k \in \{1,...,N\}, N \leq 5
\]
where $D_i$ refers to an individual code snippet, and \texttt{ResolveModule} is the function that assigns it to a specific module, $M_k$. In this step, we cluster the code snippets according to their respective modules and determine each module's priority based on its members' maximum score from the re-ranking step. In other words, a module with a single highly relevant snippet will be prioritized over one with numerous less relevant snippets. After determining the module priority, we select the top five modules for subsequent analysis.

\textbf{(b) AST-driven training loop extraction:} 
Through a comprehensive analysis of TensorFlow and PyTorch documentation, we identified eight framework-specific heuristics for detecting model training loops. They include method calls like \texttt{model.fit()}, \texttt{optimizer.step()}, and custom training loop patterns. These heuristics are available in our replication package~\cite{replicationpackage}. Using these heuristics and Abstract Syntax Tree (AST)-based parsing, we capture the potential training loops from each module:
\[
\mathcal{L}_k = \bigcup_{D_i \in M_k} \text{ExtractLoops}(\text{AST}(D_i))
\]
This extraction systematically identifies various components of a training loop, including forward/backward passes, gradient computations, optimizer steps, loss calculations, and data loader interactions. These specific components are crucial because they accurately distinguish actual training loops from other code snippets (e.g., model definitions, preprocessing scripts). The identified training loops are then used for subsequent steps.

\textbf{(c) Training loop ranking:} 
Given that a module may contain multiple training loops, we need to identify those most relevant to a reported bug. We employ the same cross-encoder\footnote{http://bit.ly/454NWFo} as the re-ranking step to assess the relevance of each training loop to the bug report, leveraging its strong understanding of both code and natural language text. Given a training loop $L_j$, and the bug report query \textit(Q), the cross-encoder computes the relevance score as follows:
\vspace{-0.3em}
\[
\text{score}_{\text{loop}}(L_j) = \text{CrossEncoder}([\text{CLS}] \oplus Q \oplus [\text{SEP}] \oplus L_j)
\]
where $\oplus$ denotes concatenation with special tokens (e.g., CLS, SEP). From each module, only one loop, specifically the one with the highest relevance score, is selected for the subsequent steps.

\textbf{(d) Construction of learning-enhanced code context:} 
Once training loops are collected, we construct the code context for every module by combining its training loop, relevant code snippets, and dependencies as follows:
\vspace{-0.5em}
\[
C_k = \underbrace{L_k}_{\text{training loop}} \cup \underbrace{\bigcup_{D_i \in \text{Top}_5(M_k)} D_i}_{\text{relevant code snippets}} \cup \underbrace{\text{DepGraph}(D_i)}_{\text{dependencies}}
\]
\[
\mathcal{C}_{\text{out}} = [C_1,...,C_K]
\]
\vspace{-0.5em}

Here, $C_k$ represents each of the top K (e.g., 5) contexts constructed.
Our reproduction agent leverages each of these contexts until it succeeds in generating the code capable of reproducing a DL bug. By combining the most relevant training loops, code snippets, and their dependencies, we thus construct a learning-enhanced code context that offers useful information to our LLM-based reproduction agent (Step 1a, Fig.~\ref{fig:schematic}).
\vspace{-0.3em}
\subsection{Bug Report Restructuring}
\looseness=-1 Using an LLM, we restructure each bug report and extract its useful information (e.g., erroneous behaviours) (Step 1b, Fig.~\ref{fig:schematic}), as follows.

First, we analyze the report title and opening paragraphs to extract the core problem statement about a bug. Second, we capture the technical details, including stack traces, error messages, performance metrics, and resource utilization patterns, to form the \emph{Observed Behaviour} section. Third, we also capture the intended system behaviour from report text, API specifications, and baseline metrics, and construct a detailed \emph{Expected Behaviour} section. Finally, RepGen synthesizes all gathered information into a detailed sequence of steps that specifies environment configurations, hardware requirements, dataset specifications, and training parameters, and constructs the steps for reproducing the bug. If any steps are explicitly mentioned in a bug report, they are leveraged by the LLM. Otherwise, the LLM reasons through the gathered information and generates a set of candidate steps. Then all three items are used by the LLM to restructure and enhance the bug report.

\subsection{Plan Generation}
\begin{figure}[!t]
\captionsetup{size = small}
\centering\includegraphics[width=0.7\columnwidth]{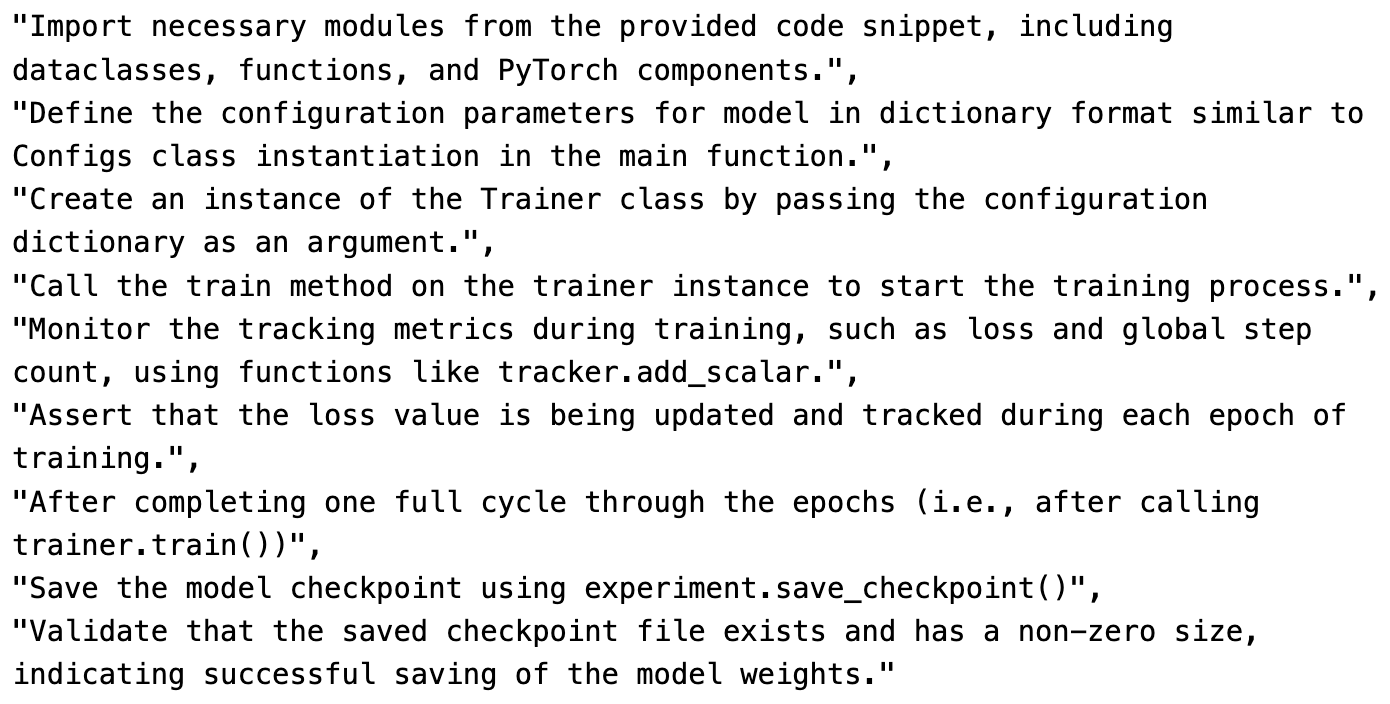}
    \vspace{-1.0em}
    \caption{Plan generated by RepGen}
    \label{fig:plan}
    \vspace{-1em}
\end{figure}

To reproduce any deep learning bug, we need a comprehensive plan. To construct the prompt for generating plans, we utilize the existing guidelines for self-planning~\cite{jiang2024planning}, which have demonstrated good performance in code generation. Self-planning has been shown in existing literature~\cite{jiang2024planning} to generate plans that produce more correct, reliable, and robust code. Using the prompt and an LLM, RepGen generates a plan that breaks the task of bug reproduction (Step 2, Fig.~\ref{fig:schematic}) into small, manageable steps. The plan guides our agent through code generation, ensuring all necessary components, configurations, and validation steps are considered. We adopt the following steps to generate our plan:

\looseness=-1
First, we identify the key components (e.g., training loops, module dependencies) and their interactions within each context using the LLM. Second, we capture the reproduction steps and numerical parameters from the bug report. Third, we structure the plan into modular, verifiable stages to enable systematic validation and control during bug reproduction attempts. In particular, it incorporates output assertions, resource monitoring, and error verification steps into the plan. Finally, we capture the plan as a structured sequence of steps to ensure seamless integration within the workflow. As a part of this step, we generate five plans (one for each context) and then pass the enriched context and its associated plan to the reproduction agent. Fig.~\ref{fig:plan} shows an example plan generated by our approach.

\subsection{Code Generation using LLM}
To execute the above plan and generate code capable of reproducing DL bugs, we introduce a specialized LLM-powered agent (Step 3, Fig.~\ref{fig:schematic}). To avoid hallucinations, our agent combines LLMs with an enriched context and a robust, multi-stage validation framework. Such a framework enables an iterative refinement process, guided by comprehensive feedback, ensuring that the generated code accurately and reliably reproduces a bug. We discuss the workflow of our bug reproduction agent below.

\looseness=-1
\textbf{Candidate Code Generation:} Our agent generates code by considering the restructured bug report, available code context, and our generated plan (Step 3a, Fig.~\ref{fig:schematic}). This initial code is likely to contain the critical components (e.g., required import statements, model architecture, training pipeline). It serves as a foundation that can be evolved to generate the final code snippet reproducing a bug.

\textbf{Capturing feedback on generated code}: To ensure syntactic correctness of our generated code, we employ AST-based \textit{structural validation} (Step 3b, Fig.~\ref{fig:schematic}). It serves as the first line of defence against syntactic errors, ensuring adherence to language grammar rules. Code that fails syntactic checks undergoes regeneration, ensuring that only syntactically sound code proceeds to subsequent stages. We employ PyLint-based \textit{static analysis} (Step 3c, Fig.~\ref{fig:schematic}) to identify potential vulnerabilities or quality issues in the syntactically correct code. It also looks for common programming errors, including missing variable definitions, incorrect import statements, invalid function calls, and type inconsistencies. The detailed feedback from this step triggers further refinement of the code by the LLM, ensuring adherence to Python best practices and minimizing potential runtime failures. Beyond syntactic correctness and completeness, we further incorporate mechanisms to ensure the relevance and effectiveness of the generated code. To avoid hallucinations and ensure relevance to the original bug report, we capture \textit{intelligent relevance feedback} against the generated code by leveraging an existing method~\cite{samir2025improvedirbasedbuglocalization} (Step 3d, Fig.~\ref{fig:schematic}). The mechanism evaluates the semantic alignment between the generated code and the bug report. If the intelligent relevance feedback indicates the irrelevance of the code, we switch to the next available context. This step is crucial given the tendency of LLMs to generate plausible but irrelevant code. Besides static feedback mechanisms above, we introduce a novel \textit{runtime feedback} mechanism that evaluates the likelihood of our generated code successfully reproducing a reported bug (Step 3e, Fig.~\ref{fig:schematic}). We capture such feedback through a systematic, multi-step process as follows. First, we approximate the final output and state of the generated code by leveraging the code understanding and reasoning capability of LLMs~\cite{chen2024reasoning}. Second, we leverage the LLM to map the program state (e.g., high loss values) to known categories of DL bugs from an established taxonomy (e.g., incorrect loss function), which we injected into the context~\cite{humbatova2020taxonomy}. If a direct match is not found in the existing taxonomy, we approximate the category of the bug based on its pre-trained knowledge. Third, the LLM leverages the taxonomy of symptoms~\cite{tambon2024silent,cao2022deepfd} provided in the prompt to derive the symptoms of the potential bugs in the code. If the symptoms are not explicitly covered in the existing taxonomy, LLM approximates the symptoms based on its pre-trained knowledge. Finally, the symptoms of the generated code are compared to those in the bug report using an LLM, which provides a similarity score. A high similarity score indicates a strong likelihood of successful bug reproduction, whereas a low score triggers an iterative refinement process. In the case of a low score, the LLM regenerates the code by incorporating the feedback on the symptoms of the code and the bug report. This dynamic feedback loop ensures continuous improvement of the generated code and its chance of reproducing the bug. Furthermore, if the agent fails to generate bug-reproducing code within five attempts, the system switches to the next available code context. Our choice of five attempts is inspired by relevant literature~\cite{wang2024aegis}. In short, our approach systematizes DL bug reproduction by delivering a comprehensive code snippet that quickly and reliably reproduces the bug.

\vspace{-2mm}
\subsection{Implementation Details}
We implement RepGen using carefully selected open-source LLMs that meet three important requirements: demonstrated excellence in code-related tasks, support for efficient quantization, and robust community maintenance. We leverage \textit{Qwen2.5-7B} for text processing tasks such as bug report restructuring, and its specialized variant \textit{Qwen2.5-Coder-7B} for code generation and analysis. They meet the above three criteria. We also adopt the recommended generation parameters~\cite{qwenparameters} for the Qwen2.5-7B and Qwen2.5-Coder-7B models in our experiment. Our hybrid retrieval framework combines both sparse and dense retrieval methods to maximize relevance. The sparse retrieval component uses BM25 (k1 = 1.2, b = 0.75) for lexical matching, while the dense retrieval component employs an efficient nearest-neighbour search (i.e., 50 trees). All experiments were conducted on machines equipped with Nvidia RTX 3050 GPUs (8GB vRAM) in a local environment. On average, each bug report was processed in $\approx$4-5 minutes, with a range between 2 and 7 minutes, depending on the complexity of the report and size of the codebase.

\vspace{-0.8em}
\section{Experiment}
We curate a benchmark dataset of 106 bugs and evaluate our technique -- RepGen -- for automated bug reproduction using standard metrics~\cite{feng2024prompting}. To contextualize our work, we compare our technique against eight state-of-the-art LLMs of varying sizes. We also analyze the impact of various components of our approach through an ablation study and investigate the cases for which our approach fails. Finally, we validate the effectiveness of our generated code by involving 27 real developers and AI engineers. In our experiments, we thus answer four research questions as follows:

\textbf{RQ1 - Effectiveness:} How effectively does RepGen reproduce deep learning bugs compared to state-of-the-art LLMs?

\textbf{RQ2 - Ablation Study:} How different components of RepGen contribute to its bug reproduction capability?

\textbf{RQ3 - Failure Analysis:} What are the current limitations of RepGen, and does increasing model size improve RepGen's ability to reproduce deep learning bugs?

\textbf{RQ4 - Usefulness:} Do developers find our generated code practically useful for bug reproduction?

\subsection{Dataset Construction}
To construct the dataset for our experiment, we followed a systematic approach inspired by prior works~\cite{defects4ml}. We select GitHub, the largest resource for open-source repositories, and use its GitHub Search API~\cite{github_api} for repository selection. We first limit our search to Python repositories due to its frequent use in DL systems~\cite{python}. Next, we identify the repositories built on three frameworks: TensorFlow, Keras, and PyTorch, the most widely used DL frameworks~\cite{frameworks}. We collected 223,485 repositories from GitHub after this step. Then, we make sure that each repository has at least one push to the main branch after January 2023, indicating their recent maintenance~\cite{defects4ml}. After this filtering, we retained 83,649 repositories for the subsequent analysis. Finally, we select such repositories that have a minimum of 10 watchers and 500 stars, a criterion used in existing literature~\cite{nikeghbal2023girt} to exclude obsolete or low-quality repositories. After applying these criteria, we collected \textbf{653} repositories from GitHub. After the preliminary filtering above, we removed repositories with fewer than 10 closed bugs, ensuring that the selected repositories have a sufficient history of reported and resolved bugs. After this filtering step, we retained \textbf{392} repositories for manual analysis. Since our primary focus was on bugs in DL systems, we manually analyzed the 392 repositories and excluded non-DL, toy or educational repositories, leading to a total of \textit{54} repositories for subsequent analysis. We then collect all the bugs that were resolved from each of the 54 repositories in the last two years (April 1 2023 - April 1 2025). We also carefully examined each bug report, fix commits, and necessary instructions and attempted to reproduce the bug in an isolated environment. If we could reproduce the bug within 1 hour, we would include this bug in the dataset. Bugs which could not be reproduced within an hour were marked as non-reproducible and excluded from further analysis. Through this process, we identified \textbf{106} reproducible bugs from \textbf{16} distinct repositories. We spent $\approx$190 person-hours on dataset construction.

\subsection{Verification of Bug Reproduction}
To verify whether our generated code accurately reproduces reported deep learning bugs, we employ a systematic approach inspired by prior work~\cite{shah2025towards}. For \textit{explicit bugs}~\cite{shah2025towards}, we confirm the bug reproduction by matching: (a) the error type (e.g., ValueError, CUDA OOM), (b) key diagnostic information (e.g., tensor shape mismatches), and (c) the execution context (e.g., training vs. inference phase). We execute our code five times with different random seeds and match their output with that of the bug report to confirm a successful reproduction. For \textit{silent bugs}~\cite{tambon2024silent}, we also execute the generated code five times with different random seeds to account for training non-determinism. The mean evaluation metrics (accuracy, loss, memory usage) from these runs are then compared against the values reported in the bug report. We consider a bug successfully reproduced if the mean metric falls within 5\% error margin of the reported value, following established practices from prior work~\cite{pham2020problems, alahmari2020challenges, shah2025towards}. We also verify the behavioural equivalence by examining whether the output of the generated code demonstrates the same failure patterns (e.g., NaN loss, excessive memory usage or performance degradation) as mentioned in the bug report.

\subsection{Baseline Selection}
To rigorously validate our approach, RepGen, we systematically chose eight open-source and closed-source LLMs as our baseline methods. We chose Llama-3 (8B, 70B), which are open-source, decoder-only transformer models with strong instruction-following capabilities and widely adopted by the research community. We also included DeepSeek-R1 (7B, 685B) and Qwen3 (8B) for their advanced reasoning capabilities. Furthermore, we included Qwen2.5-7B, as it was the base model for RepGen. Finally, we chose GPT-4.1, a highly capable proprietary model with its long context window and advanced code generation abilities. We employed three distinct prompting techniques for every LLM (zero-shot, few-shot, and chain-of-thought), and used DSPy to optimize our prompts. For smaller models (up to 8B), we conducted experiments on a local machine using Ollama, while for larger models (70B and above), we leveraged their respective APIs~\cite{openaiapi, deepseekapi, llamaapi}. We generated the code using the recommended generation parameters (e.g., temperature, top\_p, top\_k) for the local models. To mitigate the effects of LLM non-determinism, we generated the code five times and evaluated each generation following our verification protocol (Section 4.2). 

During our baseline selection, we also identified a few existing techniques. We observed that LIBRO failed to reproduce any DL bugs because it relies on high-quality, project-specific information. Other methods such as Otter(++)~\cite{ahmed2025otter} and AEGIS~\cite{wang2024aegis} were also excluded as they do not provide replication packages, making their evaluation infeasible. Similarly, as discussed earlier (Section 1), GUI-based methods do not apply to our case as well. Thus, we did not select existing bug reproduction methods for comparison due to their inapplicability to our problem scenario. 

\subsection{Developer Study}
To assess the benefits of RepGen in a real-life setting, we conduct a user study involving 27 software developers and AI engineers. We discuss our study setup, including questionnaire preparation, study design, and participant selection, as follows:

\textbf{Introduction:} In our questionnaire, we first provide necessary background context about the reproducibility of deep learning bugs and discuss the purpose of our study. Next, we provide a set of instructions and outline the structure of the subsequent sections.

\textbf{Demographic Information:} After the introduction, we gather a participant's background data through multiple-choice and numeric input questions. This includes years of professional software development experience, DL experience, and familiarity with frameworks like TensorFlow, PyTorch, Keras, and JAX. We also collect information about participants' prior experience debugging deep learning systems and the types of issues they commonly encounter.

\textbf{Bug Reproduction:} Each participant receives two bug reproduction tasks with varying difficulty levels. For each bug, participants in the experimental group received the bug report along with the generated code from RepGen, while the control group received only the bug report; the order of the reproduction tasks was randomized for each participant to prevent any ordering effects. Participants document their reproduction process, including root cause analysis, time spent, and detailed steps taken. We measure their success in bug reproduction through yes/no responses and their execution logs (if provided). We also collect qualitative data about their followed approach through open-ended questions.

\textbf{Workload Assessment:} Using NASA TLX~\cite{hart2006nasa} dimensions, we evaluate participants' experience through 5-point Likert scales that measure their mental demand, temporal demand, performance, effort, and frustration. This standardized assessment helps us quantify their cognitive load during bug reproduction.

\looseness=-1 \textbf{Code Evaluation:} The final section of our study focuses on evaluating the utility of our generated code. Through a combination of Likert scales and multiple-choice questions, participants rate the overall usefulness of the code and identify its valuable aspects for reproducibility of the bug, such as error handling, data preparation, and model configuration. Open-ended questions capture suggestions for further improvement or any complementary information.

\textbf{Participant Selection: }We first conducted a pilot study with two researchers and two developers to validate our questionnaire design. Based on their feedback, we refined the questionnaire by clarifying the ambiguous questions. We then recruited 27 professional developers and AI engineers through direct correspondence, organizational mailing lists, and professional networks. Participants were randomly assigned to either the experimental group (n = 14) or the control group (n = 13). The experimental group, which received detailed reproduction code, had a median experience of 5 years in software development and 2 years in deep learning development. Similarly, the control group also had a median of 5 years in software development and 2 years in deep learning. Both groups reported similar debugging experiences and framework expertise, with 92.57\% having prior experience in debugging DL systems, and 100\% having experience with at least one of the DL frameworks. Thus, the two groups possess equivalent relevant experience, mitigating any confounding factors in our study.

\vspace{-0.5em}
\subsection{Metrics}
To assess the effectiveness of our technique in bug reproduction, we employ the following metrics, as per the existing literature~\cite{zhao2022recdroid+, feng2024prompting}
\textbf{Success Rate (SR)} measures the percentage of deep learning bugs reproduced by a technique as follows.
$$
SR = \frac{\text{Number of successfully reproduced bugs}}{\text{Total number of bugs in the dataset}} \times 100\%
$$
\textbf{Time to Reproduce (TTR)} captures the time required to successfully reproduce a bug, measured from the start of the reproduction attempt until the bug is confirmed.

\subsection{Results}
\subsubsection{\textbf{Answering RQ$\mathbf{_1}$} -- Effectiveness}

Table~\ref{tab:baseline_results} shows the effectiveness (e.g., success rate) of RepGen and eight other baseline techniques in reproducing deep learning bugs. Our technique, RepGen, achieved an 80.19\% success rate in bug reproduction, surpassing the other baselines powered by various LLMs. While prompting has the potential to enhance an LLM's built-in reasoning abilities, it might fall short in reproducing DL bugs as follows.

First, zero-shot prompting~\cite{kojima2022large} allows LLMs to respond based on their pre-trained knowledge. They cannot access relevant codebase information, which is critical for reproducing DL bugs. As shown by Llama-3-8B's 18.87\% and Qwen2.5-7B's 21.64\% reproduction rates, zero-shot prompting results in poor performance. Even larger models, such as Llama-3-70B and GPT-4.1, struggle significantly and deliver a maximum of 49\% success rate. All these findings suggest that zero-shot prompting might not be sufficient for reproducing DL bugs. It cannot integrate project-specific context, limiting the capabilities of the LLMs. On the other hand, RepGen solves such a knowledge gap by constructing a learning-enhanced context and equipping the LLM agent with appropriate information (e.g., training pipelines, dependencies) for code generation.

\begin{table}[t]
\centering
\captionsetup{size = small}
\caption{Effectiveness of RepGen for DL Bug Reproduction}
\label{tab:baseline_results}
\setlength{\tabcolsep}{3pt}
\small
\vspace{-1.0em}
\begin{tabular}{lccc}
\toprule
\textbf{Model} & \textbf{Zero-Shot} & \textbf{Few-Shot} & \textbf{CoT} \\
\midrule
\textbf{RepGen} & \multicolumn{3}{c}{\textbf{80.19\% $\pm$1.51\%}} \\
Llama-3-8B & 18.87\% $\pm$3.29\% & 33.02\% $\pm$2.82\% & 28.30\% $\pm$2.67\% \\
Llama-3-70B & 44.34\% $\pm$2.15\% & 17.92\% $\pm$2.57\% & 17.92\% $\pm$1.41\% \\
DeepSeek-R1-7B & 36.79\% $\pm$3.66\% & 40.57\% $\pm$2.56\% & 22.64\% $\pm$3.66\% \\
DeepSeek-R1-685B & 49.06\% $\pm$1.92\% & 60.38\% $\pm$1.94\% & 42.45\% $\pm$1.64\% \\
Qwen3-8B & 23.58\% $\pm$3.07\% & 30.19\% $\pm$2.56\% & 34.91\% $\pm$3.25\% \\
GPT-4.1 & 42.45\% $\pm$1.51\% & 53.77\% $\pm$1.62\% & 40.57\% $\pm$0.75\% \\
Qwen2.5-7B & 21.70\% $\pm$2.39\% & 28.30\% $\pm$3.61\% & 30.19\% $\pm$2.06\% \\
Qwen2.5-Coder-7B & 25.47\% $\pm$3.78\% & 33.02\% $\pm$2.92\% & 28.30\% $\pm$1.94\% \\
\bottomrule
\end{tabular}
\footnotesize{\emph{Note:} RepGen does not use any in-context learning or CoT reasoning.}
\vspace{-2.5em}
\end{table}

\begin{figure}[!t]
    \centering
    \captionsetup{size = small}
\includegraphics[width=0.6\columnwidth]{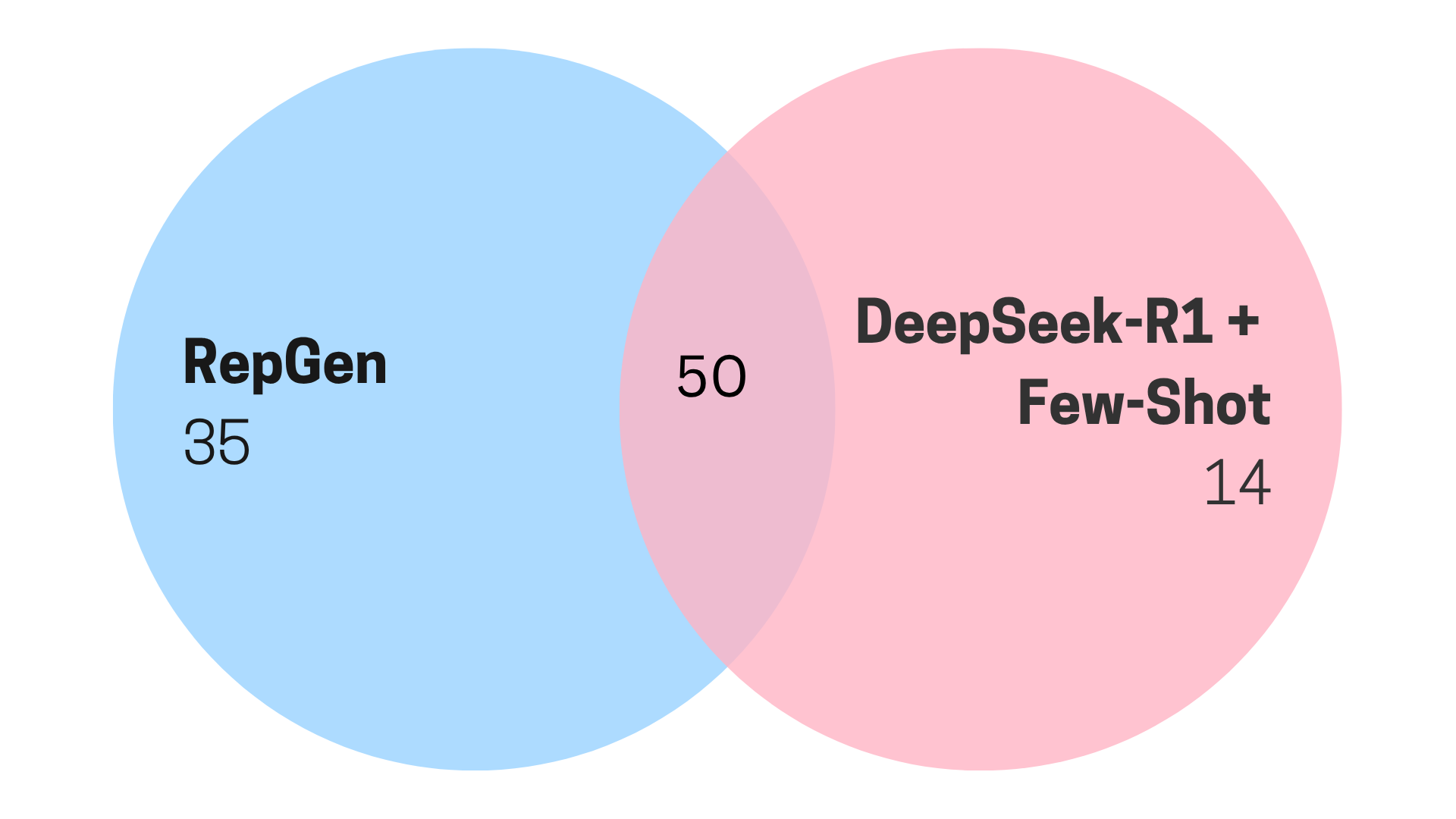}
    \vspace{-1em}
    \caption{RepGen v DeepSeek-R1}
    \label{fig:repgenvdeepseek}
    \vspace{-1.5em}
\end{figure}

Second, few-shot prompting~\cite{brown2020language} provides an LLM with a small set of example inputs and outputs to show desired behaviours and to guide the LLM's response. However, its usefulness during DL bug reproduction might be limited. While some models, like Qwen2.5-Coder-7B, show a small improvement to 33.02\% from their zero-shot performance, other models, such as Llama-3-70B, surprisingly perform worse, degrading to 17.92\% compared to their zero-shot performance. DeepSeek-R1-685B, improves significantly and reaches to 60.38\% with few-shot learning, but this is still lower than RepGen's performance. Furthermore, as shown in the Venn diagram in Fig.~\ref{fig:repgenvdeepseek}, RepGen reproduced 35 unique bugs that DeepSeek-R1 + Few-Shot could not reproduce. Thus, our findings suggest that a limited set of examples cannot possibly overcome all the challenges of DL bugs, such as the variety of bug types, non-determinism, architectural setups, and environmental settings. Furthermore, few-shot prompting is restricted by the fixed limits of the model's context window. On the other hand, RepGen overcomes these limitations by providing a dynamically constructed, enriched context tailored to specific bugs, which helps our agent, powered by the LLM, to understand the problem better and generate code capable of reproducing bugs.

Finally, chain of thought (CoT)~\cite{wei2022chain} prompting encourages LLMs to explain their intermediate reasoning steps before generating a final response. While helpful for general-purpose tasks, its use did not result in a high performance when generating code capable of reproducing deep learning bugs. Models like Llama-3-70B (17.92\%) and DeepSeek-R1-7B (22.64\%) perform worse with CoT than with zero-shot prompting. Even where small gains are seen (e.g., Qwen3-8B), the overall success in bug reproduction is still low (34.91\%). Thus, the reasoning steps of LLMs based on incomplete information or only pre-trained knowledge might not lead to an appropriate plan for a complex problem scenario like ours. More importantly, CoT has no built-in way to check if the generated code is syntactically correct, shares symptoms with a bug report, or demonstrates the erroneous runtime behaviour. On the other hand, our plan is generated by our LLM based on its reasoning capabilities, the refined bug report and the learning-enhanced context. Furthermore, we employ a multi-stage feedback mechanism, with an iterative cycle of generating, validating, and refining, which leads to code capable of reproducing the DL bugs.

We also investigated whether integrating RepGen’s artifacts enhances the reproduction capabilities of the best-performing LLM (DeepSeek-R1-685B). Using RepGen’s plan alone and context alone, DeepSeek-R1-685B reproduced 63.21\% and 68.87\% of bugs, respectively. Combining the learning-enhanced context with the plan increased this to 71.70\%. Similarly, when DeepSeek-R1-685B was provided a restructured bug report along with the context and plan, it achieved 65.04\%. This performance further improved to 72.64\% when run with an agentic setup that enhances the context, plan, and restructured report with lightweight PyLint and AST feedback.

\textbf{Statistical Significance Tests:} To demonstrate RepGen’s effectiveness, we conducted paired comparisons between RepGen and each of the 24 baseline configurations using McNemar’s test on the $2 \times 2$ contingency tables for the 106 shared bugs. McNemar’s test~\cite{mcnemar1947note} is appropriate for paired binary outcomes and for assessing whether the proportion of bugs reproduced by RepGen differs significantly from each baseline counterpart. For each comparison, we report the $p$-values and the Haldane–Anscombe adjusted odds ratio (hereafter, “odds ratio”)~\cite{haldane1956estimation}. To account for multiple comparisons, we applied the Benjamini--Hochberg correction~\cite{benjamini1995controlling} to these tests, yielding BH-adjusted $p$-values (q-values). All 24 comparisons remained statistically significant after BH correction ($q \le 3.80 \times 10^{-3}$), with odds ratios ranging from $2.45$ to $19.86$. For example, when comparing RepGen to DeepSeek-R1-685B (few-shot), there were $b = 35$ cases where RepGen succeeded and the baseline failed, versus $c = 14$ cases in the opposite direction, giving an odds ratio of $2.45$ and a BH-adjusted q-value of $3.80 \times 10^{-3}$. We also calculated 95\% confidence intervals for the difference in mean success rates for bug reproduction over five runs between RepGen and the baseline using the Newcombe hybrid score method~\cite{newcombe1998improved}, finding that the difference is $19.8\%$ with a 95\% CI of $7.0\%$--$29.7\%$, indicating that RepGen reproduces between 7\% and 30\% more of the shared bugs than DeepSeek-R1-685B. Note that this statistic differs from the odds ratio, which summarizes the relative odds of reproduction. Overall, these results demonstrate that RepGen outperforms all baselines in bug reproduction with statistical significance.

\begin{table}[t]
\captionsetup{font=small, skip=2pt}
\caption{Impact of Components on Bug Reproduction Success}
\label{tab:ablation_results_single_col_corrected}
\resizebox{0.9\columnwidth}{!}{%
\small
\centering
\begin{tabular}{|l|*{6}{c|}c|}
\hline
\multirow{2}{*}{\textbf{Config.}} & \multicolumn{6}{|c|}{\textbf{Components}} & \textbf{Success Rate} \\
\cline{2-8}
& \textbf{R1} & \textbf{R2} & \textbf{R3} & \textbf{R4} & \textbf{R5} & \textbf{R6} & \\
\hline
No Relevance F. & \tiny\ding{108} & \tiny\ding{108} & \tiny\ding{108} & \tiny\ding{108} & & \tiny\ding{108} & 17.92\% $\pm$ 1.62\% \\
\hline
No Planning & \tiny\ding{108} & & \tiny\ding{108} & \tiny\ding{108} & \tiny\ding{108} & \tiny\ding{108} & 21.70\% $\pm$ 2.18\% \\
\hline
No Restructuring & & \tiny\ding{108} & \tiny\ding{108} & \tiny\ding{108} & \tiny\ding{108} & \tiny\ding{108} & 22.64\% $\pm$ 2.18\% \\
\hline
No Compilation F. & \tiny\ding{108} & \tiny\ding{108} & & \tiny\ding{108} & \tiny\ding{108} & \tiny\ding{108} & 33.96\% $\pm$ 2.01\% \\
\hline
No Static Analysis F. & \tiny\ding{108} & \tiny\ding{108} & \tiny\ding{108} & & \tiny\ding{108} & \tiny\ding{108} & 41.51\% $\pm$ 2.15\% \\
\hline
No Runtime F. & \tiny\ding{108} & \tiny\ding{108} & \tiny\ding{108} & \tiny\ding{108} & \tiny\ding{108} & & 46.23\% $\pm$ 1.62\% \\
\hline
Complete System & \tiny\ding{108} & \tiny\ding{108} & \tiny\ding{108} & \tiny\ding{108} & \tiny\ding{108} & \tiny\ding{108} & 80.19\% $\pm$ 1.33\% \\
\hline
\end{tabular}%
}
\newline
\footnotesize\textbf{R1}=BR Restructuring, \textbf{R2}=Planning, \textbf{R3}=Compilation F., \textbf{R4}=Static Analysis F., \textbf{R5}=Relevance F., \textbf{R6}=Runtime F., \textbf{\ding{108}}=Component Present
\vspace{-1.2em}
\end{table}
\subsubsection{\textbf{Answering RQ$\mathbf{_2}$} -- Ablation Study}
Table~\ref {tab:ablation_results_single_col_corrected} shows the results of our ablation study. RepGen has six major components across different steps of its workflow. We disable each of them and assess RepGen's code generation capability towards DL bug reproduction as follows. First, we see that the removal of the relevance-checking component has the most significant impact, with performance dropping to 17.92\%. The relevance estimation between code and the bug report helps RepGen discard noisy or irrelevant code and trigger a re-generation of code. Second, the removal of the planning component reduces the success rate in bug reproduction to 21.70\%. Such a decrease indicates that breaking down bug reproduction tasks into smaller, manageable sub-tasks (a.k.a., planning) equips the LLM with a structured way to generate appropriate code and reproduce bugs. Third, the removal of the bug report restructuring component also has a major impact on RepGen and reduces the success rates to 22.64\%. This impact indicates that the restructuring of a bug report enhances its clarity and precision and thereby improves the LLM's ability to understand and reproduce issues. Fourth, the removal of compilation checks in RepGen reduced the success rate in bug reproduction to 33.96\%, underscoring the importance of syntactic correctness in generated code. Finally, the removal of static analysis feedback and runtime behaviour approximation had significant impacts, reducing success rates in bug reproduction to 46.23\% and 41.51\%, respectively. These components serve as important verification steps, ensuring that the code has relevant symptoms and is likely to reproduce the reported bugs.

\begin{table}[]
\captionsetup{font=small, skip=2pt}
\caption{Impact of Retrieval Components on Reproduction}
\label{tab:new_ablation_results}
\resizebox{0.9\columnwidth}{!}{%
\centering
\small
\begin{tabular}{|l|*{7}{c|}c|}
\hline
\multirow{2}{*}{\textbf{Config.}} & \multicolumn{7}{c|}{\textbf{Components}} & \textbf{Success Rate} \\
\cline{2-9}
& \textbf{C1} & \textbf{C2} & \textbf{C3} & \textbf{C4} & \textbf{C5} & \textbf{C6} & \textbf{C7} & \\
\hline
No ANN & & \tiny\ding{108} & \tiny\ding{108} & \tiny\ding{108} & \tiny\ding{108} & \tiny\ding{108} & \tiny\ding{108} & 26.42\% $\pm$ 2.83\% \\
\hline
No BM25 & \tiny\ding{108} & & \tiny\ding{108} & \tiny\ding{108} & \tiny\ding{108} & \tiny\ding{108} & \tiny\ding{108} & 24.53\% $\pm$ 1.89\% \\
\hline
No Dependency Ext. & \tiny\ding{108} & \tiny\ding{108} & & \tiny\ding{108} & \tiny\ding{108} & \tiny\ding{108} & \tiny\ding{108} & 34.91\% $\pm$ 0.94\% \\
\hline
No Module Part. & \tiny\ding{108} & \tiny\ding{108} & \tiny\ding{108} & & \tiny\ding{108} & \tiny\ding{108} & \tiny\ding{108} & 23.58\% $\pm$ 2.45\% \\
\hline
No Reranker & \tiny\ding{108} & \tiny\ding{108} & \tiny\ding{108} & \tiny\ding{108} & & \tiny\ding{108} & \tiny\ding{108} & 30.19\% $\pm$ 1.32\% \\
\hline
No Training Loop Ext. & \tiny\ding{108} & \tiny\ding{108} & \tiny\ding{108} & \tiny\ding{108} & \tiny\ding{108} & & \tiny\ding{108} & 28.30\% $\pm$ 2.11\% \\
\hline
No Training Loop Rank. & \tiny\ding{108} & \tiny\ding{108} & \tiny\ding{108} & \tiny\ding{108} & \tiny\ding{108} & \tiny\ding{108} & & 20.75\% $\pm$ 1.63\% \\
\hline
Complete System & \tiny\ding{108} & \tiny\ding{108} & \tiny\ding{108} & \tiny\ding{108} & \tiny\ding{108} & \tiny\ding{108} & \tiny\ding{108} & 80.19\% $\pm$ 1.33\% \\
\hline
\end{tabular}%
}
\newline
\footnotesize\textbf{C1}=ANN, \textbf{C2}=BM25, \textbf{C3}=Dependency Extraction, \textbf{C4}=Module Partitioning, \textbf{C5}=Reranker, \textbf{C6}=Training Loop Extraction, \textbf{C7}=Training Loop Ranking
\vspace{-2em}
\end{table}

We also analyze the impact of the seven components used in our context construction step (Step 1a, Fig.~\ref{fig:schematic}), with results summarized in Table~\ref{tab:new_ablation_results}. Removal of the training loop ranking component caused the largest performance drop, reducing the success rate to 20.75\%. This underscores its critical role in identifying the training loop most relevant to the reported bug, ensuring that the constructed context focuses on the correct portion of the codebase. Similarly, omitting module partitioning reduced the success rate to 23.58\%, highlighting that prioritizing modules (e.g., components or classes) based on the relevance of the code snippets rather than the sheer number of snippets is essential for building an informative context. Our hybrid retrieval mechanism also proved vital for context construction. Disabling BM25 search, which captures exact lexical matches, reduced RepGen's success rate to 24.53\%, while removing ANN search, responsible for retrieving semantically related code, resulted in a 26.42\% success rate. These drops demonstrate that both lexical and semantic retrieval are complementary and necessary for constructing a comprehensive context. Removal of the training loop extraction component lowered RepGen's performance to 28.30\%, indicating that incomplete loops without core training logic can weaken the context. Removing reranking, which refines the relevance of retrieved snippets, decreased success to 30.19\%, showing that careful prioritization of candidate code is important for context quality. Finally, omitting dependency extraction, responsible for capturing imports, helper functions, and referenced variables, reduced RepGen's success rate to 34.91\%, highlighting the need for a complete and executable context. These findings collectively show that each element of our context construction pipeline is vital for the high performance of RepGen. Our findings above align with earlier work~\cite{shah2025towards}, suggesting that successful bug reproduction requires a multi-faceted approach. The strong performance of RepGen (80.19\%) compared to its ablated versions also suggests that each component addresses a specific challenge and contributes significantly when reproducing DL bugs.
\begin{table*}[htbp]
\centering
\footnotesize
\captionsetup{size=small, skip=3pt}
\caption{Non-Reproducible Bug Categories}
\label{tab:bug_analysis_categories}
\begin{tabular}{ p{0.15\textwidth} p{0.3\textwidth} p{0.50\textwidth} }
\toprule
\textbf{Category} & \textbf{Definition} & \textbf{Example} \\
\midrule

\textbf{API/Dependency Bug} & 
An error caused by an incompatibility between the code and its dependencies (e.g., libraries), often due to version changes. &
A model fails during inference with a \texttt{TypeError} because the code passes an argument (\texttt{'end\_token\_id'}) to a function from the KerasNLP library that no longer accepts it in its updated version. \\

\textbf{Environment-Dependent Bug} & 
A bug that appears in a specific hardware or software configuration, such as a multi-GPU or distributed computing setup. &
A training job using TensorFlow's \texttt{multi\_worker\_mirrored} strategy on a two-node cluster fails to show any performance increase, indicating a scaling problem specific to the distributed environment. \\

\textbf{Data-Dependent Bug} & 
An error caused by input data issues, such as missing files, incorrect paths, or improper formatting. &
A model training script crashes during the first epoch with a \texttt{NotFoundError} because it cannot locate a file (\texttt{17e00d.jpg}) that is listed in the dataset but is not present at the path. \\
\bottomrule
\end{tabular}
\vspace{-1.3em}
\end{table*}

\subsubsection{\textbf{Answering RQ$\mathbf{_3}$} -- Failure Analysis}
To understand difficulty of reproducing certain deep learning bugs, we manually analyzed the 21 bugs that RepGen failed to reproduce. Two researchers, the lead author and an independent collaborator, each with over five years of experience in deep learning and software engineering, labelled these bugs using multiple established DL bug taxonomies~\cite{humbatova2020taxonomy, islamfse19, morovati2024bug} and identified their bug types, symptoms, and root causes. We also examined each non-reproduced bug’s report and associated developer discussion to extract its symptoms and root causes, and mapped them to the closest taxonomy-defined type. The inter-rater agreement, measured using Cohen’s kappa was 0.849, which indicates an almost perfect agreement~\cite{mchugh2012interrater}. All disagreements were subsequently resolved through discussion between the two researchers. Based on the labelled types, we then grouped the bugs into three higher-level categories: (a) API/Dependency bugs (10/21), (b) environment-dependent bugs (5/21) and (c) data-dependent bugs (6/21), as shown in Table ~\ref{tab:bug_analysis_categories}. Reproducing these categories of bugs is challenging primarily due to their reliance on specific external conditions. For example, RepGen struggles with newer APIs and updates since they were released after the knowledge cutoff of the base model, Qwen2.5-Coder-7B. Environment-dependent bugs are also difficult to reproduce because of complex distributed systems, resource allocation across multiple GPUs, and precise hardware requirements. Similarly, data-dependent bugs are challenging due to reliance on specific datasets, file system structures, and particular data formats and paths.  Moreover, we also found that each failure type aligns with distinctive pipeline signals in RepGen. API or dependency mismatches result in low BM25 or ANN relevance scores and unresolved dependency graphs. Environment-dependent bugs exhibit incomplete context coverage or unresolvable hardware-specific requirements, as indicated by runtime feedback showing that the bug cannot be reproduced in the environment created by the generated code. Data-dependent bugs produce early failures in static analysis or at runtime due to missing datasets or paths. These consistent signals explain why some bugs cannot be reproduced and provide future directions for improving RepGen's workflow.

\textbf{Experiment with GPT-4.1:} Given that our technique, empowered by Qwen2.5-Coder-7B, reproduced 85 out of 106 bugs (80.19\%), we further investigated whether a larger model could reproduce the 21 remaining bugs. We thus employed GPT-4.1, one of the most capable models, and conducted experiments using these bugs. RepGen, empowered with GPT-4.1, reproduced 8 out of the 21 bugs (38.10\%), demonstrating that larger models can tackle some of the challenging cases. This improvement is primarily due to GPT-4.1’s enhanced ability to identify and resolve API or dependency mismatches, reproducing 5 of 10 such bugs ($50.0\% \pm 15.8\%$). For instance, GPT-4.1 successfully reproduced \texttt{pip install} failures caused by invalid resolver options or \texttt{ImportError} issues stemming from incompatible Keras versions, reflecting a stronger understanding of evolving library ecosystems. Additionally, GPT-4.1’s improved code analysis enabled the reproduction of two environment-dependent bugs ($40.0\% \pm 21.9\%$) where resolution involved code modifications rather than complex infrastructure changes, such as correcting tensor dimension mismatches during training. Data-dependent bugs, such as missing image files, were largely unreproducible ($16.7\% \pm 15.2\%$) because automated techniques cannot create or manage external file system content. Conversely, RepGen, even when powered by GPT-4.1, failed to reproduce the remaining 13 bugs due to their dependence on highly specific external conditions~\cite{jahan2024towards}. Many of these were environment-dependent bugs requiring control over distributed systems, hardware configurations, or intricate framework interactions that exceed the capabilities of current LLMs. Examples include multi-node scaling issues in TensorFlow and \texttt{ModuleNotFoundError} for specialized libraries (e.g., DeepSpeed). 

\looseness=-1
Given the evidence above, model size is a factor, but not the sole determinant of success in reproducing DL bugs. This finding reinforces that the primary challenge of the automated DL bug reproduction is the lack of awareness of the bug's external context. While a larger model can resolve more issues based on its broader knowledge, it still fails when the reproduction requires precise environment-related context that it cannot access. Thus, future approaches should focus on improving the quality of the context provided to LLM agents, rather than relying on increased model sizes.

\vspace{-0.5em}
\subsubsection{\textbf{Answering RQ$\mathbf{_4}$} -- Usefulness}
To evaluate the usefulness of our generated code in a practical setting, we conducted a controlled study involving 27 professional developers and AI engineers (n=14 experimental, n=13 control). To ensure that our developer study had sufficient statistical power to detect meaningful differences between two experimental settings, we conducted an a priori power analysis. This analysis estimates whether the planned sample size can reliably detect effects of practical significance. The results indicate that our study is adequately powered (80\% power, $\alpha$=0.05) to detect large effects (Cliff's $\delta\geq$ 0.47) in continuous measures (e.g., time) and substantial differences in success rates. This aligns with typical effect sizes observed in software engineering studies on tool evaluation and is appropriate for detecting meaningful, practical improvements in bug reproduction~\cite{kampenes2007systematic, de2019evolution}. To analyze the results, we employed the same strategy as in prior work~\cite{shah2025towards}.

\vspace{-1mm}
\begin{figure}[h]
\centering
\captionsetup{size=small}
\includegraphics[width=0.85\columnwidth]{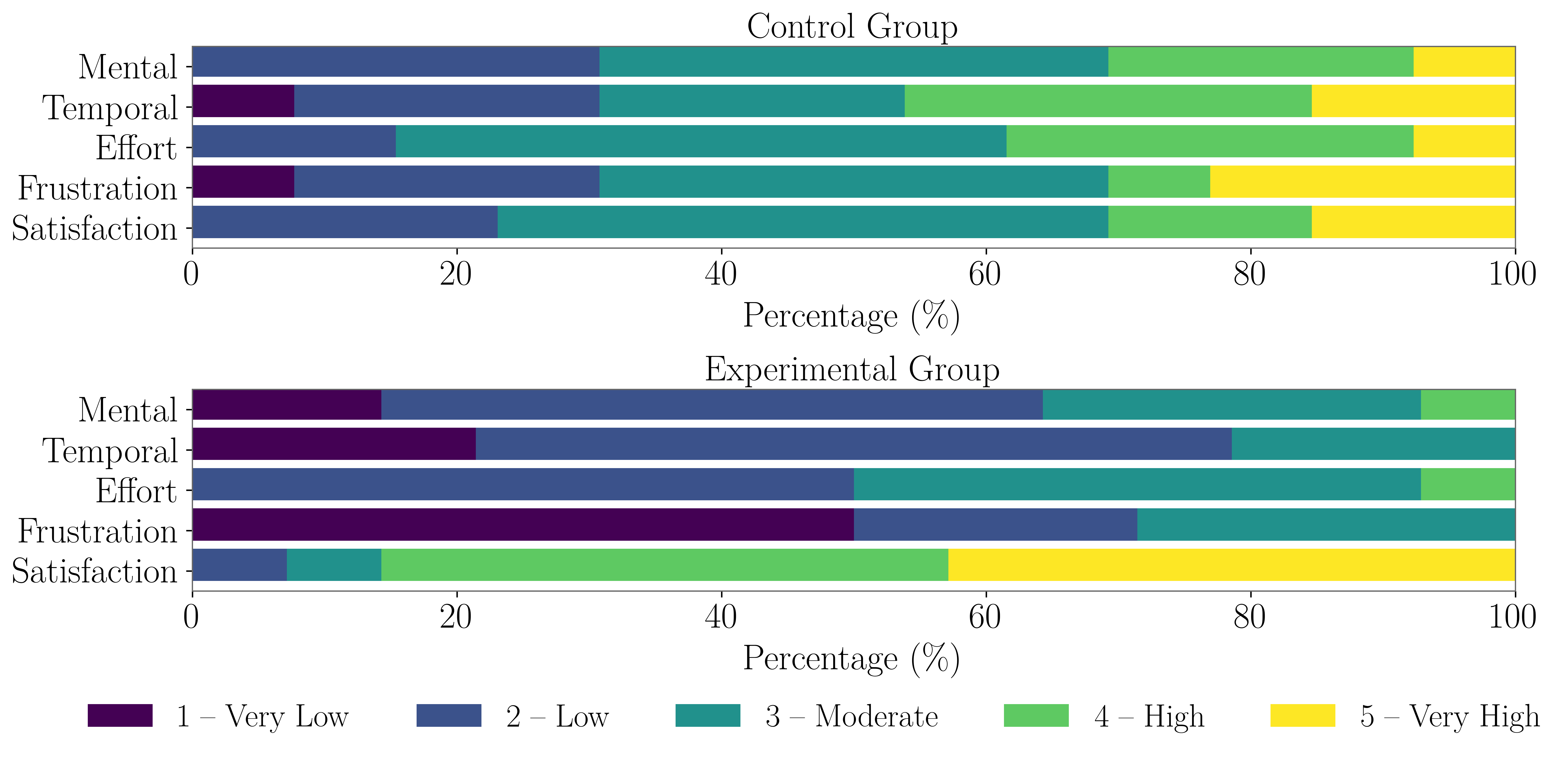}
    \vspace{-1.0em}
    \caption{Developer Study Results}         \label{fig:nasatlx}
    \vspace{-1.2em}
\end{figure}

Each participant was assigned two bug reports discussing deep learning bugs from our dataset. The control group (i.e, did not receive our generated code) achieved a 73.07\% success rate in reproducing bugs, whereas the experimental group (i.e., received our generated code) achieved a 96.42\% success rate, indicating an improvement of \textbf{23.35\% $\pm$ 18.39\%}, which show that our generated code substantially aids developers in reproducing deep learning bugs. The analysis of participants' reported reproduction process revealed that the experimental group was able to understand the bugs better. For example, one participant explained their understanding as follows: \emph{``Shape mismatch during commit loss calculation due to incorrect tensor reshaping when mask is applied''}. On the other hand, control group participants often expressed uncertainty: \emph{``Didn't understand how masking affects training''}. Our provided code also helped reduce the average time to reproduce by \textbf{56.8\% $\pm$ 16.8\%}, from 25.73 minutes for the control group to 11.11 minutes for the experimental group. For example, one participant from the experimental group explained their reproduction approach as follows: \emph{``Ran provided code, confirmed TF version, observed expected error''}. On the other hand, a participant from the control group faced challenges: \emph{``Tried multiple inputs; all failed due to unrelated tensor shape issues''}. Further details about the participants' experiences are available in our replication package~\cite{replicationpackage}.

To determine whether the presence of our generated code (i.e., categorical variable) significantly influenced the bug reproduction success of the two groups (i.e, categorical variable), we performed a $\chi^2$ test of independence. We found that there was a statistically significant difference ($p=0.0423$, $\phi=0.276$) in success rates when the code was present or absent. We also performed a Mann-Whitney $U$ test to assess the difference in the time taken by the participants from the two groups, and found out that there is a significant difference ($p=0.000$, Cliff's $\delta$ = 0.84). Thus, the above findings suggest a significant impact of RepGen in a practical setting on the success rate and time taken to reproduce deep learning bugs.

\looseness=-1 We also assessed participants' cognitive load using the NASA Task Load Index (NASA-TLX). As shown in Fig.~\ref{fig:nasatlx}, the experimental group reported significantly lower cognitive burden across all dimensions, including a 43.2\% $\pm$ 28.2\% reduction in frustration and a 25.6\% $\pm$ 23.1\% reduction in mental demand. Furthermore, participants in the experimental group also reported higher performance (4.21 vs.\ 3.23), corresponding to a {30.34\% $\pm$ 24.46\% improvement relative to the control group. Based on our collected responses, the experimental group also rated the overall usefulness of our generated code at 4.57 out of 5, with 92.30\% rating it as "Useful" or "Highly Useful" (>4). All participants in the experimental group would also prefer the inclusion of the generated code in bug reports. Thus, all the findings above suggest that RepGen's generated code offers significant benefits by improving bug-reproduction success rates, reducing time spent, and lowering developers' cognitive load.

\section{Threats To Validity}
Threats to \emph{internal validity} relate to experimental errors and biases. Replication of baseline techniques could pose a threat. To mitigate this, we used the official versions of the eight models and employed three widely used prompting techniques. We further mitigated bias by repeating our experiments five times and comparing performance across these trials~\cite{arcuri2014hitchhiker}. Threats to \emph{construct validity} concern the rigour of our evaluation methodology~\cite{smith2005construct}. During dataset construction, we implemented a stringent verification protocol involving the replication of operating environment, error manifestation, and time spent for reproduction (e.g., one-hour) to simulate real-world constraints~\cite{shah2025towards}. For silent bugs, we accepted the metrics that fall within 5\% margin of reported values, following established practices from prior work~\cite{pham2020problems, shah2025towards}. Thus, our systematic approach ensured that only verifiably reproducible bugs were included. Another threat could arise from the evaluation metrics. To mitigate this threat, we used the evaluation metrics from the existing literature~\cite{shah2025towards, feng2024prompting}. Thus, the threats to construct validity might be mitigated. Threats to \emph{external validity} relate to generalizability~\cite{findley2021external}. Our dataset contains bugs from 16 actively maintained projects spanning three frameworks, with a median of 8,282 stars and 258K LOC, indicating their strong real-world relevance. Our Python-centric implementation (e.g., PyLint integration) may warrant adaptation for non-Python DL ecosystems. However, this threat is mitigated by RepGen's modular design, which permits framework-specific adaptations. We have also released our replication package~\cite{replicationpackage} to support the reproducibility of our findings.

\section{Related Work}
\textbf{Localization and repair of deep learning bugs:} Over the last two decades, several techniques have been proposed to localize and repair bugs in deep learning (DL) systems~\cite{wardat2021deeplocalize, cao2022deepfd, zhang2021autotrainer, harzevili2024checkerbugdetectionrepair, jahan2025improved}. For example, \textsc{DeepLocalize}~\cite{wardat2021deeplocalize} uses dynamic analysis (e.g., monitoring of model parameters) to detect and localize faults in neural network training, while \textsc{DeepFD}~\cite{cao2022deepfd} applies runtime feature analysis (e.g., gradients, weights) to classify and locate faults. \textsc{AutoTrainer}~\cite{zhang2021autotrainer} monitors DL training to detect issues like vanishing gradients and applies built-in fixes. More recently, Harzevili~et~al.~\cite{harzevili2024checkerbugdetectionrepair} study input validation faults (a.k.a, checker bugs) in PyTorch and TensorFlow and leverage retrieval-augmented LLM prompts to generate their fixes. However, these methods assume the bug has already been reproduced and do not automate the reproduction of DL bugs. In contrast, we focus on automating DL bug reproduction to support downstream tasks such as fault diagnosis and repair.

\textbf{Automated bug reproduction:} Traditional approaches for bug reproduction employ GUI-based testing, program analysis, and natural language processing. For example, \textsc{ReCDroid+}~\cite{zhao2022recdroid+} synthesizes replayable event sequences from Android issue reports, whereas recent techniques like \textsc{ReBL}~\cite{feng2024prompting} and \textsc{AdbGPT}~\cite{wang2024feedback} use LLMs to extract and execute reproduction steps. These methods typically assume deterministic execution and well-defined oracles such as crashes or exceptions. Unfortunately, these techniques are not suitable for DL systems due to their non-determinism (e.g., random initialization) and complex, silent failure modes (e.g., accuracy degradation)~\cite{humbatova2020taxonomy, shah2025towards}. A recent work, \textsc{AEGIS}~\cite{wang2024aegis}, automates bug reproduction using LLM agents and external tools but overlooks DL-specific challenges such as non-determinism, hardware and dataset dependencies, weak oracles, and incomplete issue descriptions, limiting reliability. Similarly, \textsc{Otter++}~\cite{ahmed2025otter} retrieves generic code and generates test cases. In contrast, \textsc{RepGen} addresses these limitations via three key innovations: it constructs a DL-specific learning-enhanced context; integrates runtime-feedback-driven validation to estimate reproduction success; and generates complete, executable code for DL bug reproduction.

\vspace{-0.8em}
\section{Conclusion \& Future Work}
Deep learning bugs are difficult to reproduce because of the inherent nondeterminism of DL models and their tight coupling with specific hardware (e.g., GPU) and software environments (e.g., underlying frameworks). In this paper, we present \textit{RepGen}, a novel approach for automatically reproducing deep learning bugs. Our technique accelerates and systematizes the reproduction of deep learning bugs -- by integrating a learning-enhanced context, planning and a generate-validate-refine loop with multiple verification mechanisms. Evaluation on 106 real-world bugs demonstrated RepGen's superior performance, achieving an 80.19\% reproduction rate — 19.81\% higher than the best baseline. Furthermore, a controlled developer study confirmed the practical usefulness of RepGen by improving the reproduction success rates by 23.35\%, reducing average time to reproduce by 56.8\%, and substantially lowering developers' cognitive load. Future work includes extending support for distributed training bugs and integrating with existing tools localizing or repairing deep learning bugs. Thus, by enabling reliable bug reproduction, \textit{RepGen} advances the reliability and trustworthiness of DL systems.

\subsubsection*{Acknowledgments:}
This work was supported by the Natural Sciences and Engineering Research Council of Canada (Discovery Grant RGPIN-03236), the Fonds de recherche du Québec (FRQ), and the Canadian Institute for Advanced Research (CIFAR).

\balance
\bibliographystyle{ACM-Reference-Format}
\bibliography{main}


\begin{thebibliography}{84}


\ifx \showCODEN    \undefined \def \showCODEN     #1{\unskip}     \fi
\ifx \showISBNx    \undefined \def \showISBNx     #1{\unskip}     \fi
\ifx \showISBNxiii \undefined \def \showISBNxiii  #1{\unskip}     \fi
\ifx \showISSN     \undefined \def \showISSN      #1{\unskip}     \fi
\ifx \showLCCN     \undefined \def \showLCCN      #1{\unskip}     \fi
\ifx \shownote     \undefined \def \shownote      #1{#1}          \fi
\ifx \showarticletitle \undefined \def \showarticletitle #1{#1}   \fi
\ifx \showURL      \undefined \def \showURL       {\relax}        \fi
\providecommand\bibfield[2]{#2}
\providecommand\bibinfo[2]{#2}
\providecommand\natexlab[1]{#1}
\providecommand\showeprint[2][]{arXiv:#2}

\bibitem[ang(2017)]%
        {angular}
 \bibinfo{year}{2017}\natexlab{}.
\newblock \bibinfo{title}{Cosine Distance, Cosine Similarity, Angular Cosine Distance, Angular Cosine Similarity}.
\newblock
\urldef\tempurl%
\url{https://www.itl.nist.gov/div898/software/dataplot/refman2/auxillar/cosdist.htm}
\showURL{%
\tempurl}
\newblock
\shownote{Accessed: 2025-07-07}.


\bibitem[git(2022)]%
        {github_api}
 \bibinfo{year}{2022}\natexlab{}.
\newblock \bibinfo{title}{GitHub REST API Documentation}.
\newblock
\urldef\tempurl%
\url{https://docs-internal.github.com/en/rest?apiVersion=2022-11-28}
\showURL{%
\tempurl}
\newblock
\shownote{Accessed: 2025-07-02}.


\bibitem[qwe(2024)]%
        {qwenparameters}
 \bibinfo{year}{2024}\natexlab{}.
\newblock \bibinfo{title}{Generation Configuration for Qwen2.5-Coder-32B-Instruct}.
\newblock
\urldef\tempurl%
\url{https://huggingface.co/Qwen/Qwen2.5-Coder-32B-Instruct/blob/main/generation_config.json}
\showURL{%
\tempurl}
\newblock
\shownote{Accessed: 2025-07-12}.


\bibitem[fra(2024)]%
        {frameworks}
 \bibinfo{year}{2024}\natexlab{}.
\newblock \bibinfo{title}{ML Engineer comparison of Pytorch, TensorFlow, JAX, and Flax}.
\newblock
\urldef\tempurl%
\url{https://softwaremill.com/ml-engineer-comparison-of-pytorch-tensorflow-jax-and-flax/}
\showURL{%
\tempurl}
\newblock
\shownote{Accessed: 2025-06-28}.


\bibitem[pyt(2024)]%
        {python}
 \bibinfo{year}{2024}\natexlab{}.
\newblock \bibinfo{title}{What's the best programming language for machine learning?}
\newblock
\urldef\tempurl%
\url{https://www.techtarget.com/searchenterpriseai/tip/Whats-the-best-programming-language-for-machine-learning}
\showURL{%
\tempurl}
\newblock
\shownote{Accessed: 2025-07-04}.


\bibitem[dee(2025)]%
        {deepseekapi}
 \bibinfo{year}{2025}\natexlab{}.
\newblock \bibinfo{title}{DeepSeek API Documentation}.
\newblock
\urldef\tempurl%
\url{https://api-docs.deepseek.com/}
\showURL{%
\tempurl}
\newblock
\shownote{Accessed: 2025-07-17}.


\bibitem[lla(2025)]%
        {llamaapi}
 \bibinfo{year}{2025}\natexlab{}.
\newblock \bibinfo{title}{GroqCloud}.
\newblock
\urldef\tempurl%
\url{https://console.groq.com}
\showURL{%
\tempurl}
\newblock
\shownote{Accessed: 2025-07-17}.


\bibitem[ope(2025)]%
        {openaiapi}
 \bibinfo{year}{2025}\natexlab{}.
\newblock \bibinfo{title}{OpenAI API}.
\newblock
\urldef\tempurl%
\url{https://openai.com/api/}
\showURL{%
\tempurl}
\newblock
\shownote{Accessed: 2025-07-17}.


\bibitem[rep(2025)]%
        {replicationpackage}
 \bibinfo{year}{2025}\natexlab{}.
\newblock \bibinfo{title}{Replication Package for RepGen}.
\newblock
\urldef\tempurl%
\url{https://github.com/mehilshah/ICSE26-RepGen}
\showURL{%
\tempurl}


\bibitem[Addo et~al\mbox{.}(2018)]%
        {fin1}
\bibfield{author}{\bibinfo{person}{Peter~Martey Addo}, \bibinfo{person}{Dominique Guegan}, {and} \bibinfo{person}{Bertrand Hassani}.} \bibinfo{year}{2018}\natexlab{}.
\newblock \showarticletitle{Credit risk analysis using machine and deep learning models}.
\newblock \bibinfo{journal}{\emph{Risks}} \bibinfo{volume}{6}, \bibinfo{number}{2} (\bibinfo{year}{2018}), \bibinfo{pages}{38}.
\newblock


\bibitem[Ahmed et~al\mbox{.}(2025)]%
        {ahmed2025otter}
\bibfield{author}{\bibinfo{person}{Toufique Ahmed}, \bibinfo{person}{Jatin Ganhotra}, \bibinfo{person}{Rangeet Pan}, \bibinfo{person}{Avraham Shinnar}, \bibinfo{person}{Saurabh Sinha}, {and} \bibinfo{person}{Martin Hirzel}.} \bibinfo{year}{2025}\natexlab{}.
\newblock \showarticletitle{Otter: Generating Tests from Issues to Validate SWE Patches}.
\newblock \bibinfo{journal}{\emph{arXiv preprint arXiv:2502.05368}} (\bibinfo{year}{2025}).
\newblock


\bibitem[Alahmari et~al\mbox{.}(2020)]%
        {alahmari2020challenges}
\bibfield{author}{\bibinfo{person}{Saeed~S Alahmari}, \bibinfo{person}{Dmitry~B Goldgof}, \bibinfo{person}{Peter~R Mouton}, {and} \bibinfo{person}{Lawrence~O Hall}.} \bibinfo{year}{2020}\natexlab{}.
\newblock \showarticletitle{Challenges for the repeatability of deep learning models}.
\newblock \bibinfo{journal}{\emph{IEEE Access}}  \bibinfo{volume}{8} (\bibinfo{year}{2020}), \bibinfo{pages}{211860--211868}.
\newblock


\bibitem[Arcuri and Briand(2014)]%
        {arcuri2014hitchhiker}
\bibfield{author}{\bibinfo{person}{Andrea Arcuri} {and} \bibinfo{person}{Lionel Briand}.} \bibinfo{year}{2014}\natexlab{}.
\newblock \showarticletitle{A hitchhiker's guide to statistical tests for assessing randomized algorithms in software engineering}.
\newblock \bibinfo{journal}{\emph{Software Testing, Verification and Reliability}} \bibinfo{volume}{24}, \bibinfo{number}{3} (\bibinfo{year}{2014}), \bibinfo{pages}{219--250}.
\newblock


\bibitem[Bajaj et~al\mbox{.}(2016)]%
        {bajaj2016ms}
\bibfield{author}{\bibinfo{person}{Payal Bajaj}, \bibinfo{person}{Daniel Campos}, \bibinfo{person}{Nick Craswell}, \bibinfo{person}{Li Deng}, \bibinfo{person}{Jianfeng Gao}, \bibinfo{person}{Xiaodong Liu}, \bibinfo{person}{Rangan Majumder}, \bibinfo{person}{Andrew McNamara}, \bibinfo{person}{Bhaskar Mitra}, \bibinfo{person}{Tri Nguyen}, {et~al\mbox{.}}} \bibinfo{year}{2016}\natexlab{}.
\newblock \showarticletitle{Ms marco: A human generated machine reading comprehension dataset}.
\newblock \bibinfo{journal}{\emph{arXiv preprint arXiv:1611.09268}} (\bibinfo{year}{2016}).
\newblock


\bibitem[Benjamini and Hochberg(1995)]%
        {benjamini1995controlling}
\bibfield{author}{\bibinfo{person}{Yoav Benjamini} {and} \bibinfo{person}{Yosef Hochberg}.} \bibinfo{year}{1995}\natexlab{}.
\newblock \showarticletitle{Controlling the false discovery rate: a practical and powerful approach to multiple testing}.
\newblock \bibinfo{journal}{\emph{Journal of the Royal statistical society: series B (Methodological)}} \bibinfo{volume}{57}, \bibinfo{number}{1} (\bibinfo{year}{1995}), \bibinfo{pages}{289--300}.
\newblock


\bibitem[Berman et~al\mbox{.}(2019)]%
        {cyb1}
\bibfield{author}{\bibinfo{person}{Daniel~S Berman}, \bibinfo{person}{Anna~L Buczak}, \bibinfo{person}{Jeffrey~S Chavis}, {and} \bibinfo{person}{Cherita~L Corbett}.} \bibinfo{year}{2019}\natexlab{}.
\newblock \showarticletitle{A survey of deep learning methods for cyber security}.
\newblock \bibinfo{journal}{\emph{Information}} \bibinfo{volume}{10}, \bibinfo{number}{4} (\bibinfo{year}{2019}), \bibinfo{pages}{122}.
\newblock


\bibitem[Bouzenia et~al\mbox{.}(2024)]%
        {bouzenia2024repairagent}
\bibfield{author}{\bibinfo{person}{Islem Bouzenia}, \bibinfo{person}{Premkumar Devanbu}, {and} \bibinfo{person}{Michael Pradel}.} \bibinfo{year}{2024}\natexlab{}.
\newblock \showarticletitle{Repairagent: An autonomous, llm-based agent for program repair}.
\newblock \bibinfo{journal}{\emph{arXiv preprint arXiv:2403.17134}} (\bibinfo{year}{2024}).
\newblock


\bibitem[Brown et~al\mbox{.}(2020)]%
        {brown2020language}
\bibfield{author}{\bibinfo{person}{Tom Brown}, \bibinfo{person}{Benjamin Mann}, \bibinfo{person}{Nick Ryder}, \bibinfo{person}{Melanie Subbiah}, \bibinfo{person}{Jared~D Kaplan}, \bibinfo{person}{Prafulla Dhariwal}, \bibinfo{person}{Arvind Neelakantan}, \bibinfo{person}{Pranav Shyam}, \bibinfo{person}{Girish Sastry}, \bibinfo{person}{Amanda Askell}, {et~al\mbox{.}}} \bibinfo{year}{2020}\natexlab{}.
\newblock \showarticletitle{Language models are few-shot learners}.
\newblock \bibinfo{journal}{\emph{Advances in neural information processing systems}}  \bibinfo{volume}{33} (\bibinfo{year}{2020}), \bibinfo{pages}{1877--1901}.
\newblock


\bibitem[Cao et~al\mbox{.}(2022)]%
        {cao2022deepfd}
\bibfield{author}{\bibinfo{person}{Jialun Cao}, \bibinfo{person}{Meiziniu Li}, \bibinfo{person}{Xiao Chen}, \bibinfo{person}{Ming Wen}, \bibinfo{person}{Yongqiang Tian}, \bibinfo{person}{Bo Wu}, {and} \bibinfo{person}{Shing-Chi Cheung}.} \bibinfo{year}{2022}\natexlab{}.
\newblock \showarticletitle{Deepfd: Automated fault diagnosis and localization for deep learning programs}. In \bibinfo{booktitle}{\emph{Proceedings of the 44th international conference on software engineering}}. \bibinfo{pages}{573--585}.
\newblock


\bibitem[Chakraborty et~al\mbox{.}(2024)]%
        {chakraborty2024rlocator}
\bibfield{author}{\bibinfo{person}{Partha Chakraborty}, \bibinfo{person}{Mahmoud Alfadel}, {and} \bibinfo{person}{Meiyappan Nagappan}.} \bibinfo{year}{2024}\natexlab{}.
\newblock \showarticletitle{Rlocator: Reinforcement learning for bug localization}.
\newblock \bibinfo{journal}{\emph{IEEE Transactions on Software Engineering}} (\bibinfo{year}{2024}).
\newblock


\bibitem[Chen et~al\mbox{.}(2024)]%
        {chen2024reasoning}
\bibfield{author}{\bibinfo{person}{Junkai Chen}, \bibinfo{person}{Zhiyuan Pan}, \bibinfo{person}{Xing Hu}, \bibinfo{person}{Zhenhao Li}, \bibinfo{person}{Ge Li}, {and} \bibinfo{person}{Xin Xia}.} \bibinfo{year}{2024}\natexlab{}.
\newblock \showarticletitle{Reasoning Runtime Behavior of a Program with LLM: How Far Are We?}. In \bibinfo{booktitle}{\emph{2025 IEEE/ACM 47th International Conference on Software Engineering (ICSE)}}. IEEE Computer Society, \bibinfo{pages}{140--152}.
\newblock


\bibitem[Ciniselli et~al\mbox{.}(2021)]%
        {bert2021msr}
\bibfield{author}{\bibinfo{person}{Matteo Ciniselli}, \bibinfo{person}{Nathan Cooper}, \bibinfo{person}{Luca Pascarella}, \bibinfo{person}{Denys Poshyvanyk}, \bibinfo{person}{Massimiliano Di~Penta}, {and} \bibinfo{person}{Gabriele Bavota}.} \bibinfo{year}{2021}\natexlab{}.
\newblock \showarticletitle{An Empirical Study on the Usage of BERT Models for Code Completion}. In \bibinfo{booktitle}{\emph{2021 IEEE/ACM 18th International Conference on Mining Software Repositories (MSR)}}. \bibinfo{pages}{108--119}.
\newblock
\href{https://doi.org/10.1109/MSR52588.2021.00024}{doi:\nolinkurl{10.1109/MSR52588.2021.00024}}


\bibitem[de~Oliveira~Neto et~al\mbox{.}(2019)]%
        {de2019evolution}
\bibfield{author}{\bibinfo{person}{Francisco~Gomes de Oliveira~Neto}, \bibinfo{person}{Richard Torkar}, \bibinfo{person}{Robert Feldt}, \bibinfo{person}{Lucas Gren}, \bibinfo{person}{Carlo~A Furia}, {and} \bibinfo{person}{Ziwei Huang}.} \bibinfo{year}{2019}\natexlab{}.
\newblock \showarticletitle{Evolution of statistical analysis in empirical software engineering research: Current state and steps forward}.
\newblock \bibinfo{journal}{\emph{Journal of Systems and Software}}  \bibinfo{volume}{156} (\bibinfo{year}{2019}), \bibinfo{pages}{246--267}.
\newblock


\bibitem[Feng and Chen(2024)]%
        {feng2024prompting}
\bibfield{author}{\bibinfo{person}{Sidong Feng} {and} \bibinfo{person}{Chunyang Chen}.} \bibinfo{year}{2024}\natexlab{}.
\newblock \showarticletitle{Prompting is all you need: Automated android bug replay with large language models}. In \bibinfo{booktitle}{\emph{Proceedings of the 46th IEEE/ACM International Conference on Software Engineering}}. \bibinfo{pages}{1--13}.
\newblock


\bibitem[Findley et~al\mbox{.}(2021)]%
        {findley2021external}
\bibfield{author}{\bibinfo{person}{Michael~G Findley}, \bibinfo{person}{Kyosuke Kikuta}, {and} \bibinfo{person}{Michael Denly}.} \bibinfo{year}{2021}\natexlab{}.
\newblock \showarticletitle{External validity}.
\newblock \bibinfo{journal}{\emph{Annual review of political science}} \bibinfo{volume}{24}, \bibinfo{number}{1} (\bibinfo{year}{2021}), \bibinfo{pages}{365--393}.
\newblock


\bibitem[{Grand View Research}(2024)]%
        {grandview2024}
\bibfield{author}{\bibinfo{person}{{Grand View Research}}.} \bibinfo{year}{2024}\natexlab{}.
\newblock \showarticletitle{Deep Learning Market Size \& Share Report, 2024-2030}.
\newblock \bibinfo{journal}{\emph{Market Research Report}} (\bibinfo{year}{2024}).
\newblock
\urldef\tempurl%
\url{https://www.grandviewresearch.com/industry-analysis/deep-learning-market}
\showURL{%
\tempurl}


\bibitem[Grigorescu et~al\mbox{.}(2020)]%
        {grigorescu2020survey}
\bibfield{author}{\bibinfo{person}{Sorin Grigorescu}, \bibinfo{person}{Bogdan Trasnea}, \bibinfo{person}{Tiberiu Cocias}, {and} \bibinfo{person}{Gigel Macesanu}.} \bibinfo{year}{2020}\natexlab{}.
\newblock \showarticletitle{A survey of deep learning techniques for autonomous driving}.
\newblock \bibinfo{journal}{\emph{Journal of field robotics}} \bibinfo{volume}{37}, \bibinfo{number}{3} (\bibinfo{year}{2020}), \bibinfo{pages}{362--386}.
\newblock


\bibitem[Haldane(1956)]%
        {haldane1956estimation}
\bibfield{author}{\bibinfo{person}{JB Haldane}.} \bibinfo{year}{1956}\natexlab{}.
\newblock \showarticletitle{The estimation and significance of the logarithm of a ratio of frequencies}.
\newblock \bibinfo{journal}{\emph{Annals of human genetics}} \bibinfo{volume}{20}, \bibinfo{number}{4} (\bibinfo{year}{1956}), \bibinfo{pages}{309--311}.
\newblock


\bibitem[Hart(2006)]%
        {hart2006nasa}
\bibfield{author}{\bibinfo{person}{Sandra~G Hart}.} \bibinfo{year}{2006}\natexlab{}.
\newblock \showarticletitle{NASA-task load index (NASA-TLX); 20 years later}. In \bibinfo{booktitle}{\emph{Proceedings of the human factors and ergonomics society annual meeting}}, Vol.~\bibinfo{volume}{50}. Sage publications Sage CA: Los Angeles, CA, \bibinfo{pages}{904--908}.
\newblock


\bibitem[Harzevili et~al\mbox{.}(2024)]%
        {harzevili2024checkerbugdetectionrepair}
\bibfield{author}{\bibinfo{person}{Nima~Shiri Harzevili}, \bibinfo{person}{Mohammad~Mahdi Mohajer}, \bibinfo{person}{Jiho Shin}, \bibinfo{person}{Moshi Wei}, \bibinfo{person}{Gias Uddin}, \bibinfo{person}{Jinqiu Yang}, \bibinfo{person}{Junjie Wang}, \bibinfo{person}{Song Wang}, \bibinfo{person}{Zhen Ming}, \bibinfo{person}{Jiang}, {and} \bibinfo{person}{Nachiappan Nagappan}.} \bibinfo{year}{2024}\natexlab{}.
\newblock \bibinfo{title}{Checker Bug Detection and Repair in Deep Learning Libraries}.
\newblock
\showeprint[arxiv]{2410.06440}~[cs.SE]
\urldef\tempurl%
\url{https://arxiv.org/abs/2410.06440}
\showURL{%
\tempurl}


\bibitem[Honarmand and Torrellas(2014)]%
        {honarmand2014replay}
\bibfield{author}{\bibinfo{person}{Nima Honarmand} {and} \bibinfo{person}{Josep Torrellas}.} \bibinfo{year}{2014}\natexlab{}.
\newblock \showarticletitle{Replay debugging: Leveraging record and replay for program debugging}.
\newblock \bibinfo{journal}{\emph{ACM SIGARCH Computer Architecture News}} \bibinfo{volume}{42}, \bibinfo{number}{3} (\bibinfo{year}{2014}), \bibinfo{pages}{445--456}.
\newblock


\bibitem[Huang et~al\mbox{.}(2023)]%
        {huang2023demystifying}
\bibfield{author}{\bibinfo{person}{Kaifeng Huang}, \bibinfo{person}{Bihuan Chen}, \bibinfo{person}{Susheng Wu}, \bibinfo{person}{Junming Cao}, \bibinfo{person}{Lei Ma}, {and} \bibinfo{person}{Xin Peng}.} \bibinfo{year}{2023}\natexlab{}.
\newblock \showarticletitle{Demystifying dependency bugs in deep learning stack}. In \bibinfo{booktitle}{\emph{Proceedings of the 31st ACM Joint European Software Engineering Conference and Symposium on the Foundations of Software Engineering}}. \bibinfo{pages}{450--462}.
\newblock


\bibitem[Huang et~al\mbox{.}(2025)]%
        {huang2025one}
\bibfield{author}{\bibinfo{person}{Yuchao Huang}, \bibinfo{person}{Junjie Wang}, \bibinfo{person}{Zhe Liu}, \bibinfo{person}{Mingyang Li}, \bibinfo{person}{Song Wang}, \bibinfo{person}{Chunyang Chen}, \bibinfo{person}{Yuanzhe Hu}, {and} \bibinfo{person}{Qing Wang}.} \bibinfo{year}{2025}\natexlab{}.
\newblock \showarticletitle{One Sentence Can Kill the Bug: Auto-Replay Mobile App Crashes From One-Sentence Overviews}.
\newblock \bibinfo{journal}{\emph{IEEE Transactions on Software Engineering}} (\bibinfo{year}{2025}).
\newblock


\bibitem[Humbatova et~al\mbox{.}(2020)]%
        {humbatova2020taxonomy}
\bibfield{author}{\bibinfo{person}{Nargiz Humbatova}, \bibinfo{person}{Gunel Jahangirova}, \bibinfo{person}{Gabriele Bavota}, \bibinfo{person}{Vincenzo Riccio}, \bibinfo{person}{Andrea Stocco}, {and} \bibinfo{person}{Paolo Tonella}.} \bibinfo{year}{2020}\natexlab{}.
\newblock \showarticletitle{Taxonomy of real faults in deep learning systems}. In \bibinfo{booktitle}{\emph{Proceedings of the ACM/IEEE 42nd international conference on software engineering}}. \bibinfo{pages}{1110--1121}.
\newblock


\bibitem[Islam et~al\mbox{.}(2019)]%
        {islamfse19}
\bibfield{author}{\bibinfo{person}{Md~Johirul Islam}, \bibinfo{person}{Giang Nguyen}, \bibinfo{person}{Rangeet Pan}, {and} \bibinfo{person}{Hridesh Rajan}.} \bibinfo{year}{2019}\natexlab{}.
\newblock \showarticletitle{A Comprehensive Study on Deep Learning Bug Characteristics} \emph{(\bibinfo{series}{ESEC/FSE 2019})}. \bibinfo{publisher}{Association for Computing Machinery}, \bibinfo{address}{New York, NY, USA}, \bibinfo{pages}{510–520}.
\newblock
\showISBNx{9781450355728}
\href{https://doi.org/10.1145/3338906.3338955}{doi:\nolinkurl{10.1145/3338906.3338955}}


\bibitem[Jahan et~al\mbox{.}(2025)]%
        {jahan2025improved}
\bibfield{author}{\bibinfo{person}{Sigma Jahan}, \bibinfo{person}{Mehil~B Shah}, \bibinfo{person}{Parvez Mahbub}, {and} \bibinfo{person}{Mohammad~Masudur Rahman}.} \bibinfo{year}{2025}\natexlab{}.
\newblock \showarticletitle{Improved Detection and Diagnosis of Faults in Deep Neural Networks Using Hierarchical and Explainable Classification}.
\newblock \bibinfo{journal}{\emph{arXiv preprint arXiv:2501.12560}} (\bibinfo{year}{2025}).
\newblock


\bibitem[Jahan et~al\mbox{.}(2024)]%
        {jahan2024towards}
\bibfield{author}{\bibinfo{person}{Sigma Jahan}, \bibinfo{person}{Mehil~B Shah}, {and} \bibinfo{person}{Mohammad~Masudur Rahman}.} \bibinfo{year}{2024}\natexlab{}.
\newblock \showarticletitle{Towards understanding the challenges of bug localization in deep learning systems}.
\newblock \bibinfo{journal}{\emph{arXiv preprint arXiv:2402.01021}} (\bibinfo{year}{2024}).
\newblock


\bibitem[Jiang et~al\mbox{.}(2024)]%
        {jiang2024planning}
\bibfield{author}{\bibinfo{person}{Xue Jiang}, \bibinfo{person}{Yihong Dong}, \bibinfo{person}{Lecheng Wang}, \bibinfo{person}{Zheng Fang}, \bibinfo{person}{Qiwei Shang}, \bibinfo{person}{Ge Li}, \bibinfo{person}{Zhi Jin}, {and} \bibinfo{person}{Wenpin Jiao}.} \bibinfo{year}{2024}\natexlab{}.
\newblock \showarticletitle{Self-Planning Code Generation with Large Language Models}.
\newblock \bibinfo{journal}{\emph{ACM Trans. Softw. Eng. Methodol.}} \bibinfo{volume}{33}, \bibinfo{number}{7}, Article \bibinfo{articleno}{182} (\bibinfo{date}{Sept.} \bibinfo{year}{2024}), \bibinfo{numpages}{30}~pages.
\newblock
\showISSN{1049-331X}
\href{https://doi.org/10.1145/3672456}{doi:\nolinkurl{10.1145/3672456}}


\bibitem[Kampenes et~al\mbox{.}(2007)]%
        {kampenes2007systematic}
\bibfield{author}{\bibinfo{person}{Vigdis~By Kampenes}, \bibinfo{person}{Tore Dyb{\aa}}, \bibinfo{person}{Jo~E Hannay}, {and} \bibinfo{person}{Dag~IK Sj{\o}berg}.} \bibinfo{year}{2007}\natexlab{}.
\newblock \showarticletitle{A systematic review of effect size in software engineering experiments}.
\newblock \bibinfo{journal}{\emph{Information and Software Technology}} \bibinfo{volume}{49}, \bibinfo{number}{11-12} (\bibinfo{year}{2007}), \bibinfo{pages}{1073--1086}.
\newblock


\bibitem[Kang et~al\mbox{.}(2023)]%
        {kang2023large}
\bibfield{author}{\bibinfo{person}{Sungmin Kang}, \bibinfo{person}{Juyeon Yoon}, {and} \bibinfo{person}{Shin Yoo}.} \bibinfo{year}{2023}\natexlab{}.
\newblock \showarticletitle{Large language models are few-shot testers: Exploring llm-based general bug reproduction}. In \bibinfo{booktitle}{\emph{2023 IEEE/ACM 45th International Conference on Software Engineering (ICSE)}}. IEEE, \bibinfo{pages}{2312--2323}.
\newblock


\bibitem[Karmakar and Robbes(2021)]%
        {karmakar2021pre}
\bibfield{author}{\bibinfo{person}{Anjan Karmakar} {and} \bibinfo{person}{Romain Robbes}.} \bibinfo{year}{2021}\natexlab{}.
\newblock \showarticletitle{What do pre-trained code models know about code?}. In \bibinfo{booktitle}{\emph{2021 36th IEEE/ACM International Conference on Automated Software Engineering (ASE)}}. IEEE, \bibinfo{pages}{1332--1336}.
\newblock


\bibitem[Kojima et~al\mbox{.}(2022)]%
        {kojima2022large}
\bibfield{author}{\bibinfo{person}{Takeshi Kojima}, \bibinfo{person}{Shixiang~Shane Gu}, \bibinfo{person}{Machel Reid}, \bibinfo{person}{Yutaka Matsuo}, {and} \bibinfo{person}{Yusuke Iwasawa}.} \bibinfo{year}{2022}\natexlab{}.
\newblock \showarticletitle{Large language models are zero-shot reasoners}.
\newblock \bibinfo{journal}{\emph{Advances in neural information processing systems}}  \bibinfo{volume}{35} (\bibinfo{year}{2022}), \bibinfo{pages}{22199--22213}.
\newblock


\bibitem[Lewis et~al\mbox{.}(2020)]%
        {lewis2020retrieval}
\bibfield{author}{\bibinfo{person}{Patrick Lewis}, \bibinfo{person}{Ethan Perez}, \bibinfo{person}{Aleksandra Piktus}, \bibinfo{person}{Fabio Petroni}, \bibinfo{person}{Vladimir Karpukhin}, \bibinfo{person}{Naman Goyal}, \bibinfo{person}{Heinrich K{\"u}ttler}, \bibinfo{person}{Mike Lewis}, \bibinfo{person}{Wen-tau Yih}, \bibinfo{person}{Tim Rockt{\"a}schel}, {et~al\mbox{.}}} \bibinfo{year}{2020}\natexlab{}.
\newblock \showarticletitle{Retrieval-augmented generation for knowledge-intensive nlp tasks}.
\newblock \bibinfo{journal}{\emph{Advances in neural information processing systems}}  \bibinfo{volume}{33} (\bibinfo{year}{2020}), \bibinfo{pages}{9459--9474}.
\newblock


\bibitem[Li et~al\mbox{.}(2022)]%
        {li2022dear}
\bibfield{author}{\bibinfo{person}{Yi Li}, \bibinfo{person}{Shaohua Wang}, {and} \bibinfo{person}{Tien~N Nguyen}.} \bibinfo{year}{2022}\natexlab{}.
\newblock \showarticletitle{Dear: A novel deep learning-based approach for automated program repair}. In \bibinfo{booktitle}{\emph{Proceedings of the 44th international conference on software engineering}}. \bibinfo{pages}{511--523}.
\newblock


\bibitem[Mahbub and Rahman(2024)]%
        {mahbub2024predicting}
\bibfield{author}{\bibinfo{person}{Parvez Mahbub} {and} \bibinfo{person}{Mohammad~Masudur Rahman}.} \bibinfo{year}{2024}\natexlab{}.
\newblock \showarticletitle{Predicting line-level defects by capturing code contexts with hierarchical transformers}. In \bibinfo{booktitle}{\emph{2024 IEEE International Conference on Software Analysis, Evolution and Reengineering (SANER)}}. IEEE, \bibinfo{pages}{308--319}.
\newblock


\bibitem[Mahbub et~al\mbox{.}(2023)]%
        {mahbub2023explaining}
\bibfield{author}{\bibinfo{person}{Parvez Mahbub}, \bibinfo{person}{Ohiduzzaman Shuvo}, {and} \bibinfo{person}{Mohammad~Masudur Rahman}.} \bibinfo{year}{2023}\natexlab{}.
\newblock \showarticletitle{Explaining software bugs leveraging code structures in neural machine translation}. In \bibinfo{booktitle}{\emph{2023 IEEE/ACM 45th International Conference on Software Engineering (ICSE)}}. IEEE, \bibinfo{pages}{640--652}.
\newblock


\bibitem[Mahmud et~al\mbox{.}(2024)]%
        {mahmud2024using}
\bibfield{author}{\bibinfo{person}{Junayed Mahmud}, \bibinfo{person}{Nadeeshan De~Silva}, \bibinfo{person}{Safwat~Ali Khan}, \bibinfo{person}{Seyed~Hooman Mostafavi}, \bibinfo{person}{SM~Hasan Mansur}, \bibinfo{person}{Oscar Chaparro}, \bibinfo{person}{Andrian Marcus}, {and} \bibinfo{person}{Kevin Moran}.} \bibinfo{year}{2024}\natexlab{}.
\newblock \showarticletitle{On using gui interaction data to improve text retrieval-based bug localization}. In \bibinfo{booktitle}{\emph{Proceedings of the 46th IEEE/ACM International Conference on Software Engineering}}. \bibinfo{pages}{1--13}.
\newblock


\bibitem[McHugh(2012)]%
        {mchugh2012interrater}
\bibfield{author}{\bibinfo{person}{Mary~L McHugh}.} \bibinfo{year}{2012}\natexlab{}.
\newblock \showarticletitle{Interrater reliability: the kappa statistic}.
\newblock \bibinfo{journal}{\emph{Biochemia medica}} \bibinfo{volume}{22}, \bibinfo{number}{3} (\bibinfo{year}{2012}), \bibinfo{pages}{276--282}.
\newblock


\bibitem[McNemar(1947)]%
        {mcnemar1947note}
\bibfield{author}{\bibinfo{person}{Quinn McNemar}.} \bibinfo{year}{1947}\natexlab{}.
\newblock \showarticletitle{Note on the sampling error of the difference between correlated proportions or percentages}.
\newblock \bibinfo{journal}{\emph{Psychometrika}} \bibinfo{volume}{12}, \bibinfo{number}{2} (\bibinfo{year}{1947}), \bibinfo{pages}{153--157}.
\newblock


\bibitem[Morovati et~al\mbox{.}(2023)]%
        {defects4ml}
\bibfield{author}{\bibinfo{person}{Mohammad~Mehdi Morovati}, \bibinfo{person}{Amin Nikanjam}, \bibinfo{person}{Foutse Khomh}, {and} \bibinfo{person}{Zhen Ming~(Jack) Jiang}.} \bibinfo{year}{2023}\natexlab{}.
\newblock \showarticletitle{Bugs in Machine Learning-Based Systems: A Faultload Benchmark}.
\newblock \bibinfo{journal}{\emph{Empirical Softw. Engg.}} \bibinfo{volume}{28}, \bibinfo{number}{3} (\bibinfo{date}{apr} \bibinfo{year}{2023}), \bibinfo{numpages}{33}~pages.
\newblock
\showISSN{1382-3256}
\href{https://doi.org/10.1007/s10664-023-10291-1}{doi:\nolinkurl{10.1007/s10664-023-10291-1}}


\bibitem[Morovati et~al\mbox{.}(2024)]%
        {morovati2024bug}
\bibfield{author}{\bibinfo{person}{Mohammad~Mehdi Morovati}, \bibinfo{person}{Amin Nikanjam}, \bibinfo{person}{Florian Tambon}, \bibinfo{person}{Foutse Khomh}, {and} \bibinfo{person}{Zhen~Ming Jiang}.} \bibinfo{year}{2024}\natexlab{}.
\newblock \showarticletitle{Bug characterization in machine learning-based systems}.
\newblock \bibinfo{journal}{\emph{Empirical Software Engineering}} \bibinfo{volume}{29}, \bibinfo{number}{1} (\bibinfo{year}{2024}), \bibinfo{pages}{14}.
\newblock


\bibitem[Nagarajan et~al\mbox{.}(2018)]%
        {nagarajan2018impact}
\bibfield{author}{\bibinfo{person}{Prabhat Nagarajan}, \bibinfo{person}{Garrett Warnell}, {and} \bibinfo{person}{Peter Stone}.} \bibinfo{year}{2018}\natexlab{}.
\newblock \showarticletitle{The impact of nondeterminism on reproducibility in deep reinforcement learning}.
\newblock  (\bibinfo{year}{2018}).
\newblock


\bibitem[Nayrolles et~al\mbox{.}(2015)]%
        {nayrolles2015jcharming}
\bibfield{author}{\bibinfo{person}{Mathieu Nayrolles}, \bibinfo{person}{Abdelwahab Hamou-Lhadj}, \bibinfo{person}{Sofi{\`e}ne Tahar}, {and} \bibinfo{person}{Alf Larsson}.} \bibinfo{year}{2015}\natexlab{}.
\newblock \showarticletitle{JCHARMING: A bug reproduction approach using crash traces and directed model checking}. In \bibinfo{booktitle}{\emph{2015 IEEE 22nd International Conference on Software Analysis, Evolution, and Reengineering (SANER)}}. IEEE, \bibinfo{pages}{101--110}.
\newblock


\bibitem[Naziri et~al\mbox{.}(2025)]%
        {naziri2025bugsindlls}
\bibfield{author}{\bibinfo{person}{MM~Abid Naziri}, \bibinfo{person}{Aman~Kumar Singh}, \bibinfo{person}{Benjamin Wu}, \bibinfo{person}{Feiran Qin}, \bibinfo{person}{Saikat Dutta}, {and} \bibinfo{person}{Marcelo d'Amorim}.} \bibinfo{year}{2025}\natexlab{}.
\newblock \showarticletitle{BugsInDLLs: A Database of Reproducible Bugs in Deep Learning Libraries to Enable Systematic Evaluation of Testing Techniques}. In \bibinfo{booktitle}{\emph{Proceedings of the 34th ACM SIGSOFT International Symposium on Software Testing and Analysis}}. \bibinfo{pages}{61--65}.
\newblock


\bibitem[Newcombe(1998)]%
        {newcombe1998improved}
\bibfield{author}{\bibinfo{person}{Robert~G Newcombe}.} \bibinfo{year}{1998}\natexlab{}.
\newblock \showarticletitle{Improved confidence intervals for the difference between binomial proportions based on paired data}.
\newblock \bibinfo{journal}{\emph{Statistics in medicine}} \bibinfo{volume}{17}, \bibinfo{number}{22} (\bibinfo{year}{1998}), \bibinfo{pages}{2635--2650}.
\newblock


\bibitem[Nikeghbal et~al\mbox{.}(2023)]%
        {nikeghbal2023girt}
\bibfield{author}{\bibinfo{person}{Nafiseh Nikeghbal}, \bibinfo{person}{Amir~Hossein Kargaran}, \bibinfo{person}{Abbas Heydarnoori}, {and} \bibinfo{person}{Hinrich Sch{\"u}tze}.} \bibinfo{year}{2023}\natexlab{}.
\newblock \showarticletitle{Girt-data: Sampling github issue report templates}. In \bibinfo{booktitle}{\emph{2023 IEEE/ACM 20th International Conference on Mining Software Repositories (MSR)}}. IEEE, \bibinfo{pages}{104--108}.
\newblock


\bibitem[Pham et~al\mbox{.}(2020)]%
        {pham2020problems}
\bibfield{author}{\bibinfo{person}{Hung~Viet Pham}, \bibinfo{person}{Shangshu Qian}, \bibinfo{person}{Jiannan Wang}, \bibinfo{person}{Thibaud Lutellier}, \bibinfo{person}{Jonathan Rosenthal}, \bibinfo{person}{Lin Tan}, \bibinfo{person}{Yaoliang Yu}, {and} \bibinfo{person}{Nachiappan Nagappan}.} \bibinfo{year}{2020}\natexlab{}.
\newblock \showarticletitle{Problems and opportunities in training deep learning software systems: An analysis of variance}. In \bibinfo{booktitle}{\emph{Proceedings of the 35th IEEE/ACM international conference on automated software engineering}}. \bibinfo{pages}{771--783}.
\newblock


\bibitem[{PwC}(2023)]%
        {pwc2023}
\bibfield{author}{\bibinfo{person}{{PwC}}.} \bibinfo{year}{2023}\natexlab{}.
\newblock \bibinfo{booktitle}{\emph{AI Predictions Survey}}.
\newblock \bibinfo{type}{{T}echnical {R}eport}. \bibinfo{institution}{PricewaterhouseCoopers}.
\newblock
\urldef\tempurl%
\url{https://www.pwc.com/us/en/tech-effect/ai-analytics/ai-predictions.html}
\showURL{%
\tempurl}


\bibitem[Robertson et~al\mbox{.}(2009)]%
        {robertson2009probabilistic}
\bibfield{author}{\bibinfo{person}{Stephen Robertson}, \bibinfo{person}{Hugo Zaragoza}, {et~al\mbox{.}}} \bibinfo{year}{2009}\natexlab{}.
\newblock \showarticletitle{The probabilistic relevance framework: BM25 and beyond}.
\newblock \bibinfo{journal}{\emph{Foundations and Trends{\textregistered} in Information Retrieval}} \bibinfo{volume}{3}, \bibinfo{number}{4} (\bibinfo{year}{2009}), \bibinfo{pages}{333--389}.
\newblock


\bibitem[Saha et~al\mbox{.}(2013)]%
        {saha2013improving}
\bibfield{author}{\bibinfo{person}{Ripon~K Saha}, \bibinfo{person}{Matthew Lease}, \bibinfo{person}{Sarfraz Khurshid}, {and} \bibinfo{person}{Dewayne~E Perry}.} \bibinfo{year}{2013}\natexlab{}.
\newblock \showarticletitle{Improving bug localization using structured information retrieval}. In \bibinfo{booktitle}{\emph{2013 28th IEEE/ACM International Conference on Automated Software Engineering (ASE)}}. IEEE, \bibinfo{pages}{345--355}.
\newblock


\bibitem[Samir and Rahman(2025)]%
        {samir2025improvedirbasedbuglocalization}
\bibfield{author}{\bibinfo{person}{Asif~Mohammed Samir} {and} \bibinfo{person}{Mohammad~Masudur Rahman}.} \bibinfo{year}{2025}\natexlab{}.
\newblock \bibinfo{title}{Improved IR-based Bug Localization with Intelligent Relevance Feedback}.
\newblock
\showeprint[arxiv]{2501.10542}~[cs.SE]
\urldef\tempurl%
\url{https://arxiv.org/abs/2501.10542}
\showURL{%
\tempurl}


\bibitem[Shah et~al\mbox{.}(2024)]%
        {shah2024towards}
\bibfield{author}{\bibinfo{person}{Mehil~B Shah}, \bibinfo{person}{Mohammad~Masudur Rahman}, {and} \bibinfo{person}{Foutse Khomh}.} \bibinfo{year}{2024}\natexlab{}.
\newblock \showarticletitle{Towards Understanding the Impact of Data Bugs on Deep Learning Models in Software Engineering}.
\newblock \bibinfo{journal}{\emph{arXiv preprint arXiv:2411.12137}} (\bibinfo{year}{2024}).
\newblock


\bibitem[Shah et~al\mbox{.}(2025)]%
        {shah2025towards}
\bibfield{author}{\bibinfo{person}{Mehil~B Shah}, \bibinfo{person}{Mohammad~Masudur Rahman}, {and} \bibinfo{person}{Foutse Khomh}.} \bibinfo{year}{2025}\natexlab{}.
\newblock \showarticletitle{Towards enhancing the reproducibility of deep learning bugs: an empirical study}.
\newblock \bibinfo{journal}{\emph{Empirical Software Engineering}} \bibinfo{volume}{30}, \bibinfo{number}{1} (\bibinfo{year}{2025}), \bibinfo{pages}{23}.
\newblock


\bibitem[Shen et~al\mbox{.}(2017)]%
        {med1}
\bibfield{author}{\bibinfo{person}{Dinggang Shen}, \bibinfo{person}{Guorong Wu}, {and} \bibinfo{person}{Heung-Il Suk}.} \bibinfo{year}{2017}\natexlab{}.
\newblock \showarticletitle{Deep learning in medical image analysis}.
\newblock \bibinfo{journal}{\emph{Annual review of biomedical engineering}}  \bibinfo{volume}{19} (\bibinfo{year}{2017}), \bibinfo{pages}{221--248}.
\newblock


\bibitem[Smith(2005)]%
        {smith2005construct}
\bibfield{author}{\bibinfo{person}{Gregory~T Smith}.} \bibinfo{year}{2005}\natexlab{}.
\newblock \showarticletitle{On construct validity: issues of method and measurement.}
\newblock \bibinfo{journal}{\emph{Psychological assessment}} \bibinfo{volume}{17}, \bibinfo{number}{4} (\bibinfo{year}{2005}), \bibinfo{pages}{396}.
\newblock


\bibitem[Soltani et~al\mbox{.}(2018)]%
        {soltani2018single}
\bibfield{author}{\bibinfo{person}{Mozhan Soltani}, \bibinfo{person}{Pouria Derakhshanfar}, \bibinfo{person}{Annibale Panichella}, \bibinfo{person}{Xavier Devroey}, \bibinfo{person}{Andy Zaidman}, {and} \bibinfo{person}{Arie van Deursen}.} \bibinfo{year}{2018}\natexlab{}.
\newblock \showarticletitle{Single-objective versus multi-objectivized optimization for evolutionary crash reproduction}. In \bibinfo{booktitle}{\emph{Search-Based Software Engineering: 10th International Symposium, SSBSE 2018, Montpellier, France, September 8-9, 2018, Proceedings 10}}. Springer, \bibinfo{pages}{325--340}.
\newblock


\bibitem[Tambon et~al\mbox{.}(2024)]%
        {tambon2024silent}
\bibfield{author}{\bibinfo{person}{Florian Tambon}, \bibinfo{person}{Amin Nikanjam}, \bibinfo{person}{Le An}, \bibinfo{person}{Foutse Khomh}, {and} \bibinfo{person}{Giuliano Antoniol}.} \bibinfo{year}{2024}\natexlab{}.
\newblock \showarticletitle{Silent bugs in deep learning frameworks: an empirical study of keras and tensorflow}.
\newblock \bibinfo{journal}{\emph{Empirical Software Engineering}} \bibinfo{volume}{29}, \bibinfo{number}{1} (\bibinfo{year}{2024}), \bibinfo{pages}{10}.
\newblock


\bibitem[Tymchuk et~al\mbox{.}(2018)]%
        {tymchuk2018jit}
\bibfield{author}{\bibinfo{person}{Yuriy Tymchuk}, \bibinfo{person}{Mohammad Ghafari}, {and} \bibinfo{person}{Oscar Nierstrasz}.} \bibinfo{year}{2018}\natexlab{}.
\newblock \showarticletitle{JIT feedback: What experienced developers like about static analysis}. In \bibinfo{booktitle}{\emph{Proceedings of the 26th Conference on Program Comprehension}}. \bibinfo{pages}{64--73}.
\newblock


\bibitem[Wakabayashi(2018)]%
        {selfdrivingcarcrash}
\bibfield{author}{\bibinfo{person}{Daisuke Wakabayashi}.} \bibinfo{year}{2018}\natexlab{}.
\newblock \bibinfo{title}{Self-driving uber car kills pedestrian in Arizona, where Robots Roam}.
\newblock
\urldef\tempurl%
\url{https://www.nytimes.com/2018/03/19/technology/uber-driverless-fatality.html}
\showURL{%
\tempurl}
\newblock
\shownote{Accessed on December 17, 2023}.


\bibitem[Wang et~al\mbox{.}(2025)]%
        {wang2025llmdroid}
\bibfield{author}{\bibinfo{person}{Chenxu Wang}, \bibinfo{person}{Tianming Liu}, \bibinfo{person}{Yanjie Zhao}, \bibinfo{person}{Minghui Yang}, {and} \bibinfo{person}{Haoyu Wang}.} \bibinfo{year}{2025}\natexlab{}.
\newblock \showarticletitle{LLMDroid: Enhancing Automated Mobile App GUI Testing Coverage with Large Language Model Guidance}.
\newblock \bibinfo{journal}{\emph{Proceedings of the ACM on Software Engineering}} \bibinfo{volume}{2}, \bibinfo{number}{FSE} (\bibinfo{year}{2025}), \bibinfo{pages}{1001--1022}.
\newblock


\bibitem[Wang et~al\mbox{.}(2024a)]%
        {wang2024systematic}
\bibfield{author}{\bibinfo{person}{Di Wang}, \bibinfo{person}{Matthias Galster}, {and} \bibinfo{person}{Miguel Morales-Trujillo}.} \bibinfo{year}{2024}\natexlab{a}.
\newblock \showarticletitle{A systematic mapping study of bug reproduction and localization}.
\newblock \bibinfo{journal}{\emph{Information and Software Technology}}  \bibinfo{volume}{165} (\bibinfo{year}{2024}), \bibinfo{pages}{107338}.
\newblock


\bibitem[Wang et~al\mbox{.}(2022a)]%
        {wang2022bridging}
\bibfield{author}{\bibinfo{person}{Deze Wang}, \bibinfo{person}{Zhouyang Jia}, \bibinfo{person}{Shanshan Li}, \bibinfo{person}{Yue Yu}, \bibinfo{person}{Yun Xiong}, \bibinfo{person}{Wei Dong}, {and} \bibinfo{person}{Xiangke Liao}.} \bibinfo{year}{2022}\natexlab{a}.
\newblock \showarticletitle{Bridging pre-trained models and downstream tasks for source code understanding}. In \bibinfo{booktitle}{\emph{Proceedings of the 44th international conference on software engineering}}. \bibinfo{pages}{287--298}.
\newblock


\bibitem[Wang et~al\mbox{.}(2024c)]%
        {wang2024feedback}
\bibfield{author}{\bibinfo{person}{Dingbang Wang}, \bibinfo{person}{Yu Zhao}, \bibinfo{person}{Sidong Feng}, \bibinfo{person}{Zhaoxu Zhang}, \bibinfo{person}{William~GJ Halfond}, \bibinfo{person}{Chunyang Chen}, \bibinfo{person}{Xiaoxia Sun}, \bibinfo{person}{Jiangfan Shi}, {and} \bibinfo{person}{Tingting Yu}.} \bibinfo{year}{2024}\natexlab{c}.
\newblock \showarticletitle{Feedback-driven automated whole bug report reproduction for android apps}. In \bibinfo{booktitle}{\emph{Proceedings of the 33rd ACM SIGSOFT International Symposium on Software Testing and Analysis}}. \bibinfo{pages}{1048--1060}.
\newblock


\bibitem[Wang et~al\mbox{.}(2024b)]%
        {wang2024aegis}
\bibfield{author}{\bibinfo{person}{Xinchen Wang}, \bibinfo{person}{Pengfei Gao}, \bibinfo{person}{Xiangxin Meng}, \bibinfo{person}{Ruida Hu}, \bibinfo{person}{Chao Peng}, \bibinfo{person}{Yun Lin}, {and} \bibinfo{person}{Cuiyun Gao}.} \bibinfo{year}{2024}\natexlab{b}.
\newblock \showarticletitle{AEGIS: An Agent-based Framework for Bug Reproduction from Issue Descriptions}.
\newblock  (\bibinfo{year}{2024}).
\newblock


\bibitem[Wang et~al\mbox{.}(2022b)]%
        {wang-etal-2022-compilable}
\bibfield{author}{\bibinfo{person}{Xin Wang}, \bibinfo{person}{Yasheng Wang}, \bibinfo{person}{Yao Wan}, \bibinfo{person}{Fei Mi}, \bibinfo{person}{Yitong Li}, \bibinfo{person}{Pingyi Zhou}, \bibinfo{person}{Jin Liu}, \bibinfo{person}{Hao Wu}, \bibinfo{person}{Xin Jiang}, {and} \bibinfo{person}{Qun Liu}.} \bibinfo{year}{2022}\natexlab{b}.
\newblock \showarticletitle{Compilable Neural Code Generation with Compiler Feedback}. In \bibinfo{booktitle}{\emph{Findings of the Association for Computational Linguistics: ACL 2022}}, \bibfield{editor}{\bibinfo{person}{Smaranda Muresan}, \bibinfo{person}{Preslav Nakov}, {and} \bibinfo{person}{Aline Villavicencio}} (Eds.). \bibinfo{publisher}{Association for Computational Linguistics}, \bibinfo{address}{Dublin, Ireland}, \bibinfo{pages}{9--19}.
\newblock
\href{https://doi.org/10.18653/v1/2022.findings-acl.2}{doi:\nolinkurl{10.18653/v1/2022.findings-acl.2}}


\bibitem[Wardat et~al\mbox{.}(2021)]%
        {wardat2021deeplocalize}
\bibfield{author}{\bibinfo{person}{Mohammad Wardat}, \bibinfo{person}{Wei Le}, {and} \bibinfo{person}{Hridesh Rajan}.} \bibinfo{year}{2021}\natexlab{}.
\newblock \showarticletitle{Deeplocalize: Fault localization for deep neural networks}. In \bibinfo{booktitle}{\emph{2021 IEEE/ACM 43rd International Conference on Software Engineering (ICSE)}}. IEEE, \bibinfo{pages}{251--262}.
\newblock


\bibitem[Wei et~al\mbox{.}(2022)]%
        {wei2022chain}
\bibfield{author}{\bibinfo{person}{Jason Wei}, \bibinfo{person}{Xuezhi Wang}, \bibinfo{person}{Dale Schuurmans}, \bibinfo{person}{Maarten Bosma}, \bibinfo{person}{Fei Xia}, \bibinfo{person}{Ed Chi}, \bibinfo{person}{Quoc~V Le}, \bibinfo{person}{Denny Zhou}, {et~al\mbox{.}}} \bibinfo{year}{2022}\natexlab{}.
\newblock \showarticletitle{Chain-of-thought prompting elicits reasoning in large language models}.
\newblock \bibinfo{journal}{\emph{Advances in neural information processing systems}}  \bibinfo{volume}{35} (\bibinfo{year}{2022}), \bibinfo{pages}{24824--24837}.
\newblock


\bibitem[Xia et~al\mbox{.}(2023)]%
        {xia2023automated}
\bibfield{author}{\bibinfo{person}{Chunqiu~Steven Xia}, \bibinfo{person}{Yuxiang Wei}, {and} \bibinfo{person}{Lingming Zhang}.} \bibinfo{year}{2023}\natexlab{}.
\newblock \showarticletitle{Automated program repair in the era of large pre-trained language models}. In \bibinfo{booktitle}{\emph{2023 IEEE/ACM 45th International Conference on Software Engineering (ICSE)}}. IEEE, \bibinfo{pages}{1482--1494}.
\newblock


\bibitem[Zhang et~al\mbox{.}(2024)]%
        {zhang2024ragenhancedcommitmessagegeneration}
\bibfield{author}{\bibinfo{person}{Linghao Zhang}, \bibinfo{person}{Hongyi Zhang}, \bibinfo{person}{Chong Wang}, {and} \bibinfo{person}{Peng Liang}.} \bibinfo{year}{2024}\natexlab{}.
\newblock \bibinfo{title}{RAG-Enhanced Commit Message Generation}.
\newblock
\showeprint[arxiv]{2406.05514}~[cs.SE]
\urldef\tempurl%
\url{https://arxiv.org/abs/2406.05514}
\showURL{%
\tempurl}


\bibitem[Zhang et~al\mbox{.}(2021)]%
        {zhang2021autotrainer}
\bibfield{author}{\bibinfo{person}{Xiaoyu Zhang}, \bibinfo{person}{Juan Zhai}, \bibinfo{person}{Shiqing Ma}, {and} \bibinfo{person}{Chao Shen}.} \bibinfo{year}{2021}\natexlab{}.
\newblock \showarticletitle{Autotrainer: An automatic dnn training problem detection and repair system}. In \bibinfo{booktitle}{\emph{2021 IEEE/ACM 43rd International Conference on Software Engineering (ICSE)}}. IEEE, \bibinfo{pages}{359--371}.
\newblock


\bibitem[Zhang et~al\mbox{.}(2018)]%
        {tensorflowprogrambugs}
\bibfield{author}{\bibinfo{person}{Yuhao Zhang}, \bibinfo{person}{Yifan Chen}, \bibinfo{person}{Shing-Chi Cheung}, \bibinfo{person}{Yingfei Xiong}, {and} \bibinfo{person}{Lu Zhang}.} \bibinfo{year}{2018}\natexlab{}.
\newblock \showarticletitle{An empirical study on tensorflow program bugs}. In \bibinfo{booktitle}{\emph{ISSTA}}. \bibinfo{pages}{129--140}.
\newblock


\bibitem[Zhang et~al\mbox{.}(2020)]%
        {zhang2020detecting}
\bibfield{author}{\bibinfo{person}{Yuhao Zhang}, \bibinfo{person}{Luyao Ren}, \bibinfo{person}{Liqian Chen}, \bibinfo{person}{Yingfei Xiong}, \bibinfo{person}{Shing-Chi Cheung}, {and} \bibinfo{person}{Tao Xie}.} \bibinfo{year}{2020}\natexlab{}.
\newblock \showarticletitle{Detecting numerical bugs in neural network architectures}. In \bibinfo{booktitle}{\emph{Proceedings of the 28th ACM joint meeting on european software engineering conference and symposium on the foundations of software engineering}}. \bibinfo{pages}{826--837}.
\newblock


\bibitem[Zhao et~al\mbox{.}(2022)]%
        {zhao2022recdroid+}
\bibfield{author}{\bibinfo{person}{Yu Zhao}, \bibinfo{person}{Ting Su}, \bibinfo{person}{Yang Liu}, \bibinfo{person}{Wei Zheng}, \bibinfo{person}{Xiaoxue Wu}, \bibinfo{person}{Ramakanth Kavuluru}, \bibinfo{person}{William~GJ Halfond}, {and} \bibinfo{person}{Tingting Yu}.} \bibinfo{year}{2022}\natexlab{}.
\newblock \showarticletitle{Recdroid+: Automated end-to-end crash reproduction from bug reports for android apps}.
\newblock \bibinfo{journal}{\emph{ACM Transactions on Software Engineering and Methodology (TOSEM)}} \bibinfo{volume}{31}, \bibinfo{number}{3} (\bibinfo{year}{2022}), \bibinfo{pages}{1--33}.
\newblock


\bibitem[Zhao et~al\mbox{.}(2019)]%
        {zhao2019recdroid}
\bibfield{author}{\bibinfo{person}{Yu Zhao}, \bibinfo{person}{Tingting Yu}, \bibinfo{person}{Ting Su}, \bibinfo{person}{Yang Liu}, \bibinfo{person}{Wei Zheng}, \bibinfo{person}{Jingzhi Zhang}, {and} \bibinfo{person}{William~GJ Halfond}.} \bibinfo{year}{2019}\natexlab{}.
\newblock \showarticletitle{Recdroid: automatically reproducing android application crashes from bug reports}. In \bibinfo{booktitle}{\emph{2019 IEEE/ACM 41st International Conference on Software Engineering (ICSE)}}. IEEE, \bibinfo{pages}{128--139}.
\newblock


\end{thebibliography}
\end{document}